\documentclass[useAMS,usenatbib]{mnras}

\usepackage{tabto}
\usepackage{float}
\usepackage{color}
\usepackage{amssymb} 
\usepackage{amsfonts,amsmath,amscd} 
\usepackage[subnum]{cases}
\usepackage{mathtools}
\usepackage{verbatim}
\usepackage{t1enc}
\usepackage[T1]{fontenc}
\usepackage{ae,aecompl}
\usepackage[dvips]{graphics}
\usepackage{adjustbox}
\usepackage{eufrak} 
\usepackage{euscript} 
\usepackage{graphicx} 
\usepackage[normalem]{ulem} 
\usepackage{makeidx} 
\usepackage[usenames,dvipsnames]{pstricks}
\usepackage{epsfig}
\usepackage{pst-grad} 
\usepackage{pst-plot} 
\usepackage{pstcol,times,graphics,pstricks,pst-node,pst-coil} 
\usepackage{multicol} 
\usepackage{multirow} 
\usepackage{indentfirst} 
\usepackage{natbib} 
\usepackage{setspace} 
\usepackage{threeparttable}

\title[CVs in MOCCA GCs -- progenitors and CV formation]
{MOCCA-SURVEY database I. Accreting white dwarf binary systems in globular clusters -- II. Cataclysmic variables -- progenitors and population at birth}
\author[Belloni et al.]{Diogo Belloni$^{1,2}$\thanks{E-mail: belloni@camk.edu.pl (DB)}
Mirek Giersz$^1$\thanks{E-mail: mig@camk.edu.pl (MG)}, Helio J. Rocha-Pinto$^{3}$, Nathan W. C. Leigh$^{4}$, 
\newauthor Abbas Askar $^{1}$\\
$^{1}$ Nicolaus Copernicus Astronomical Centre, Polish Academy of Sciences, ul. Bartycka 18, PL-00-716 Warsaw, Poland \\
$^{2}$ CAPES Foundation, Ministry of Education of Brazil, DF 70040-020, Brasilia, Brazil \\
$^{3}$ Universidade Federal do Rio de Janeiro, Observat{\'o}rio do Valongo, Ladeira do Pedro Ant\^onio 43, 20080-090 Rio de Janeiro, Brazil \\
$^{4}$ Department of Astrophysics, American Museum of Natural History, Central Park West and 79th Street, New York, NY 10024, USA}
\begin{document}

\date{Accepted 2016 September 30. Received 2016 September 28; in original form 2016 August 26}

\pagerange{\pageref{firstpage}--\pageref{lastpage}} \pubyear{2016}

\maketitle

\label{firstpage}


\begin{abstract}
This is the second in a series of papers associated with cataclysmic
variables (CVs) and related objects, formed in a suite of simulations for globular cluster 
evolution performed with the MOCCA Monte Carlo code.  We study the 
properties of our simulated CV populations throughout the entire cluster evolution. 
We find that dynamics extends the range of binary CV progenitor properties,
causing CV formation from binary progenitors that would otherwise not become CVs.  
The CV formation rate in our simulations can be separated into two regimes: an initial burst 
($\lesssim$ 1 Gyr) connected with the formation of the most massive white dwarfs, followed by a 
nearly constant formation rate.  This result holds for all models regardless of the adopted 
initial conditions, even when most CVs form dynamically.  Given the cluster age-dependence 
of CV properties, we argue that direct comparisons to observed Galactic field CVs could 
be misleading, since cluster CVs can be up to 4 times older than their field counterparts.  Our 
results also illustrate that, due mainly to unstable mass transfer, some CVs that 
form in our simulations are destroyed before the present-day.  Finally, some field CVs might 
have originated from globular clusters, as found in our simulations, although the fraction of such 
escapers should be small relative to the entire Galactic field CV population.
\end{abstract}


\section{INTRODUCTION}

Cataclysmic variables (CVs) are interacting
binaries composed of a white dwarf (WD) undergoing stable 
mass transfer from a main sequence (MS) star or a brown dwarf (BD) 
\citep[e.g.][]{Warner_1995_OK,Knigge_2011_OK}.  They are 
expected to exist in non-negligible numbers in globular clusters
(GCs), that are natural laboratories for testing theories of stellar 
dynamics and evolution. 

CVs in GCs have been studied by many authors, both 
theoretically and observationally 
\citep[e.g.][and references therein]{Knigge_2012MMSAI}.  GCs 
are thought to play a crucial role in CV formation, since their 
densities are sufficiently high that dynamical encounters 
involving binaries should be common.  Thus, in dense 
GCs, it is natural to expect that many CV progenitors will have 
been affected by dynamics in some way prior to CV formation 
\citep[e.g.][]{Ivanova_2006,Belloni_2016a}. 

\subsection{Formation and Destruction Channels}
\label{formation_channel}

The primary channels associated with CV formation in GCs, 
based on the results of numerical simulations, can be summarized 
as \citep[][]{Ivanova_2006} following:
(i) only $\sim$ 27 per cent of CVs form from binaries that have 
never experienced a dynamical interaction;
(ii) $\sim$ 60 per cent of CVs did not evolve via a common-envelope
phase (CEP);
(iii) tidal capture does not play a significant role in CV formation; and
(iv) dynamical encounters tend to exchange more massive
WDs into the binary progenitors of CVs.

Some of these formation channels have been independently
discussed by \citet{Shara_2006}. The authors analyzed
two cluster $N$-body models, simulated using the NBODY4 code 
with GRAPE-6 processors. In their simulations, they found 4 (out of 15)
CVs had no field-like counterpart, i.e. four CVs were dynamically
formed, either by exchange or successive dynamical
encounters.

In both \citet{Ivanova_2006} and \citet{Shara_2006}, the authors also 
discussed destruction channels for any CVs that do not survive
to the present-day in their simulations.  \citet{Ivanova_2006} found
that most CVs cease mass transfer for (internal) 
evolutionary reasons; very few CVs are destroyed 
by dynamical encounters. \citet{Shara_2006} found
similar results in this regard, however fewer CVs managed
to escape from their simulated clusters relative to what was found by 
\citet{Ivanova_2006}, and the mechanism of escape was 
always weak encounters or two-body relaxation.

In spite of the agreement between these two modeling efforts, we emphasize 
that \citet{Shara_2006} analyzed only two models (100k and 200k) and 
had very few CVs form (< 20), whereas \citet{Ivanova_2006} analyzed several 
Monte Carlo models with many CVs, but lacked any cluster evolution, i.e.
their clusters had fixed spatial structure.  We aim to complement these 
pioneering works with the MOCCA Monte Carlo simulations presented here, 
which offer both statistical significance and a realistic dynamical environment 
for the host clusters.

\subsection{Dynamical Influence on CV formation and evolution}
\label{dynamical_cv_evolution}

One interesting feature described by \citet{Shara_2006} is a dynamically-induced 
acceleration in the onset and rate of accretion when CVs are formed.  For example, 
the authors discuss a particular CV formed in their simulations that began 
the CV phase before its field-like counterpart, due to a prior dynamical 
interaction.  Another CV had its evolution accelerated by a dynamical perturbation 
just after initiating the CV phase.

We argue that the first of these two scenarios is more likely, since 
CV progenitors tend to be more massive than the mean stellar and even binary mass in 
the cluster, which reduces the time-scale for dynamical encounters 
prior to the onset of the CV phase. If these interactions decrease the binary 
progenitor orbital period (or increase its eccentricity), then the CE phase will 
begin earlier than for its field counterpart.  This scenario would not necessarily rely on 
a single strong interaction, but could rather arise due to repeated weak interactions.  

The second scenario, on the other hand, seems
much less likely, since this requires (very few) strong dynamical interactions and 
hence small impact parameters.  CVs tend to have short orbital 
periods, which reduces the probability for direct encounters with 
sufficient force to significantly change their orbital parameters \citep{Leigh_2016}.  However,
if the CV is formed close to or in the cluster core, then
the probability of a strong interaction is at its highest.  
This was the case for the specific CV that suffered accelerated mass transfer 
due to a dynamical perturbation in the simulations of \citet{Shara_2006}, since this 
binary was very close to the core at its formation.

We further caution the reader that these conclusions taken from \citet{Shara_2006} 
rely on small number CV statistics, albeit accompanied by rich details for their formation 
and subsequent time evolution.  Hence, we emphasize that these results need 
to be confirmed, by supplanting the small number statistics with a more robust 
coverage of the relevant parameter space for GC evolution and CV formation, using 
a much larger suite of realistic simulations.

\subsection{CV age}
\label{cv_age}

We emphasize that observing CVs in GCs relies on much more than just 
dynamical interactions.  Of comparable or even greater importance 
are the observational selection effects, as well as 
the ages of cluster CVs compared to Galactic field CVs.
For instance, \citet{Belloni_2016a} quantified the observational 
selection effects that plague the search for CVs in GCs, and concluded that
their detection rates could be dramatically increased if detectable during 
quiescence.  \citet{Ak_2015} inferred the ages of a 
sample of field CVs from kinematic data, and concluded that 94 per cent of CVs
in the solar neighbourhood belong to the thin-disc component 
of the Galaxy.  The corresponding mean kinematical ages are 3.40 $\pm$ 1.03 Gyr 
and 3.90 $\pm$ 1.28 Gyr for the non-magnetic thin-disc CVs below and above 
the period gap, respectively.  In GCs, on the other hand, some CVs can be up to 4 
times older than this.  Thus, it is critical to properly account for such age-related 
effects when comparing cluster and field CV populations, in an attempt to quantify 
the impact of the cluster dynamics on CV formation and evolution.

\subsection{Structure of the paper}
\label{paper_structure}

For clarity, we have separated the results of our initial investigation into CV formation 
in GCs into two different papers.  In the first paper in this series \citep[][]{Belloni_2016a},
we concentrated on the present-day population (PDP) of CVs and the observational selection
effects that contribute to "hiding" most of the CV populations in GCs from observations.

In this paper, the second of the series, we focus on the primary formation
channels for CVs, as simulated by the MOCCA code (Section \ref{mocca}),
and quantify the influence of the cluster dynamics in shaping the observed CV properties.  
We further address the age-dependence of CV properties, CVs destroyed before the 
present-day (i.e., after 12 Gyr of cluster evolution) and CVs formed from binary progenitors that 
previously escaped their host cluster.

In Section \ref{model}, we describe the MOCCA and CATUABA codes and 
present the suite of models analyzed in this paper. In Section \ref{results}, 
the main results of this investigation are presented and discussed.  We 
conclude and summarize our main results in Section \ref{conclusion}.

\section{METHODOLOGY AND MODELS}
\label{model}

In this section, we briefly describe the 
MOCCA code that was used to simulate the six cluster realizations considered here, 
the CATUABA code that was used to analyze the simulated CV populations in each 
model and, finally, the initial cluster conditions.

\subsection{MOCCA code}
\label{mocca}

The MOCCA code is based on the orbit-averaged Monte Carlo technique 
for cluster evolution developed by \citet{Henon_1971}, which 
was further improved by  \citet{Stodolkiewicz_1986}, and then 
developed even further by \citet[][and references therein]{Giersz_2013}.  These 
last authors included in MOCCA the FEWBODY code, developed by 
\citet{Fregeau_2004} to perform numerical scattering experiments of 
small-number gravitational interactions.  
To model the Galactic potential, MOCCA assumes a point-mass with 
total mass equal to the enclosed Galaxy mass at the specified Galactocentric 
radius. The description of escape processes in tidally limited 
clusters follows the procedure derived by 
\citet{Fukushige_2000}. The stellar evolution is implemented via
 the SSE code developed by \citet{Hurley_2000} for single stars and the 
 BSE code developed by \citet{Hurley_2002} for binary evolution. 

The most important part of MOCCA with regards to CV formation and related
exotic objects is the BSE code \citep{Hurley_2002}. Here, as in 
\citet[][]{Belloni_2016a}, we adopt the standard parameters
described in \citet{Hurley_2002}, which includes the adoption
of their assumed efficiency parameter for the CE phase 
($\alpha =$ 3)\footnote{BSE assumes that the common envelope
binding energy is that of the giant(s) envelope involved
in the process. In order to reconcile the prescription
developed by \citet{Iben_Livio_1993}, the recommended value for 
the CEP efficiency is 3.0.}. We emphasize that the defaults in BSE, 
and especially the assumption for $\alpha$, may be over-estimates of 
realistic values \citep[e.g.][]{Zorotovic_2010,Camacho_2014}. However, 
our focus in this paper is to quantify the role of cluster dynamics 
and evolution on CV properties in globular clusters.  We defer a more 
thorough investigation of the various aspects of binary evolution relevant 
for CV formation to a forthcoming paper.  For now, we note simply that our 
adopted assumptions for the BSE code should mainly affect the total numbers 
of CVs formed via binary evolution (in particular the CEP).  Our conclusions 
regarding the influence of the cluster dynamics on CV formation should 
be roughly insensitive to these assumptions.

MOCCA has been extensively tested against N-body codes. 
For instance, \citet{Giersz_2013} showed that MOCCA 
reproduces the results of $N$-body codes with excellent precision, including not 
only the rate of cluster evolution and the subsequent re-structuring of 
the cluster mass distribution, but also the calculated distributions of binary orbital 
parameters.  Additionally, \citet{Wang_2016} compared MOCCA with
the state-of-the-art NBODY6++GPU, showing good agreement
between the predictions of the two codes.

To summarize, due to its fast, efficient and accurate coverage of the relevant parameter space, 
MOCCA is ideal for performing big surveys aimed at modeling large populations of CVs in 
many GCs \citep[][]{Leigh_2013,Leigh_2015,Giersz_2015}, and for studying in detail the influence of the host 
cluster environment on the properties 
of different types of exotic objects such as CVs \citep[][this work]{Belloni_2016a}, blue straggler stars 
\citep[][]{Hypki_2013,Hypki_2016a,Hypki_2016b}, 
intermediate-mass black holes \citep[][]{Giersz_2015}, X-ray binaries, etc.

\subsection{CATUABA code}
\label{catuaba}

The CATUABA ({\bf C}ode for {\bf A}nalyzing and 
s{\bf TU}dying c{\bf A}taclysmic variables, sym{\bf B}iotic stars
and {\bf A}M CVns) code was introduced and described in \citet{Belloni_2016a}.  
The code was created for the purpose of analyzing populations
of CVs and related objects (AM CVns and symbiotic stars)
produced in clusters simulated by the MOCCA code.

In its current version, it allows for a detailed analysis of the entire 
CV population produced over the course of the simulated 
cluster evolution.  It combines the BSE code with the disc instability
model (DIM) constrained by empirical data (in order to fill in any gaps 
in binary evolution not accounted for in BSE) to calculate the observed 
CV properties. 

CATUABA permits for the unambiguous identification of a given CV population
in the cluster snapshots provided by MOCCA at various times in the cluster evolution.  
It then compiles the most relevant events (related to both stellar and binary evolution, 
as well as dynamical interactions) in the history of all individual CVs.  Armed with this 
information, CATUABA unequivocally separates the CVs into three groups, defined by 
the influence of dynamical interactions incurred over the course of the binary progenitor 
lifetimes.  The first group contains only CVs formed without any influence from dynamics 
(BSE group), the second group includes CVs weakly influenced by dynamics (WDI group), 
and the last group includes those CVs strongly influenced by dynamical interactions 
(SDI group).

Some comments are worth mentioning here regarding the distinction between the WDI and 
SDI groups \citep[][see section 3.2 for more details]{Belloni_2016a}:

\begin{enumerate}
\item Whether a CV belongs to the WDI group or the SDI group 
is a priori arbitrary.  We define this transition based on the 
fraction of energy exchanged between the target binary and 
interloping single or binary star at the end of a dynamical 
interaction, relative to the initial orbital energy of the 
target binary.  The exact cutoff between weak and strong 
dynamical interactions is 20 per cent of the initial binding 
energy.  We emphasize that this cutoff applies only on a per 
interaction basis, and is an arbitrarily chosen value based 
on the average energy exchange given in Spitzer's formula for 
equal-mass systems \citep{Spitzer_BOOK}. Generally, for the WDI 
group, most of the interactions are fly-bys.  However, resonant 
interactions do sometimes occur, although relatively rarely by 
comparison. For the SDI group, most interactions are resonant 
and/or exchanges.  According to our definition, a binary experiences 
an WDI if it undergoes a single interaction and incurs a change in 
its binding energy that is less than 20 per cent of its initial value.  
Otherwise, the binary was subjected to an SDI. If a given binary 
experiences only weak dynamical interactions over the course of its 
lifetime, then this binary belongs to the WDI group.  Conversely, if 
the binary underwent one or more strong dynamical interactions over 
the course of its lifetime, then this binary belongs to the SDI group.

\item The WDI group is formed by CVs that had only weak 
encounters during their lives. However, if the number
of these encounters is high for a CV in this group, 
their cumulative effect can be strong and the properties
of the progenitor binary can be strongly changed during its life.
Otherwise, if the cumulative effect of the weak dynamical
interactions is not strong, the CV will probably have
similar properties to those belonging to the BSE group.

\item Only the SDI group includes binaries that
underwent either exchange or merger.
\end{enumerate}

The CATUABA code also provides information regarding the progenitor binary 
populations, and the formation-age populations. The progenitor population is defined
as the population of all binaries that are CV progenitors, 
i.e. the population of `CVs' at the time of cluster birth (i.e., $t=0$).  The formation-age 
population is defined as the population of binary progenitors at the time of initiation of the 
CV phase (i.e., when mass transfer starts from the donor on to the WD).  Obviously, for the
formation-age population, this time is not unique, since the time corresponding to the onset of 
the CV phase is different for every CV.  

Finally, CATUABA allows for the study of CVs that form in the cluster
but do not survive up to the present-day (i.e., destroyed CVs), while also for the clear 
identification of CVs formed from escaped binaries (for ease of comparison to 
analogous field CV populations formed from primordial binaries born in the field).  
In short, CATUABA identifies, organizes and quantifies the time evolution of CV properties 
in realistic evolving GCs.  

All features and main assumptions concerning the CATUABA code are explained in
more detail in \citet[][see their section 3]{Belloni_2016a}.

\subsection{Models}
\label{models}

In this work, we continue the analysis started 
by \citet{Belloni_2016a}, who investigated models with two distinct
initial binary populations (IBPs).  Each IBP contains all initial binaries in a given
cluster, and is associated with a specific set of binary orbital parameter distributions.  
Briefly, the six models analyzed in \citet{Belloni_2016a} differ mainly 
with respect to the initial central density, the initial binary fraction and 
choice of the IBP. 

The first set of models, defined as the K (or Kroupa) models,
correspond to models with 95 per cent primordial binaries, with 
the IBPs having binary orbital parameter distributions that follow the 
Kroupa IBP \citep{Kroupa_INITIAL}, which is derived from eigenevolution 
and mass feeding algorithms \citep{Kroupa_INITIAL}.
 
The second set of models, defined as the S (or Standard) models, correspond to models with 
5 or 10 per cent primordial binaries, with the IBPs having binary orbital parameter distributions that 
follow the Standard IBP.  
The Standard IBP follows a 
uniform distribution for the binary mass ratio, an uniform distribution in the 
logarithm of the semi-major axis $\log(a)$ or a log-normal semi-major axis 
distribution, and a thermal eccentricity distribution.  For these two sets 
of models, we consider sparse ($\rho_c \sim 10^3$ M$_\odot$ pc$^{-3}$), 
dense ($\rho_c \sim 10^5$ M$_\odot$ pc$^{-3}$), 
and very dense ($\rho_c > 10^5$ M$_\odot$ pc$^{-3}$) clusters,
where $\rho_c$ is the cluster central density.

We assume that all stars are on the zero-age main sequence
when the simulation begins, and that all residual gas remaining from the star
formation process has already been removed from the cluster at t $=$ 0.  
All models adopt a low metallicity ($\lesssim$ 0.001), are initially in virial 
equilibrium, and have neither rotation nor primordial mass segregation.  Finally, 
all models are evolved up to a cluster age of 12 Gyr, which is taken to be the 
"present-day".

The properties of all initial models are summarized in Table
1 of \citet{Belloni_2016a}.  We have also evolved these same models (actually, just 
the IBP) with BSE alone over a 12 Gyr period.  In this way, we create a "control" sample of 
CVs, for comparison to the cluster CVs formed in our Monte Carlo simulations.  This 
control sample is therefore tailored for comparison to observed CV populations in the 
Galactic field.  This will help us to disentangle the role of the cluster dynamics in shaping the 
properties of CV populations in GCs, for all six initial models considered here.

After describing the main codes performing the required calculations and the six sets of 
initial conditions considered, in the next section we turn our attention to the results of our analysis 
of these models.  We begin by presenting our main results concerning the progenitor (i.e., initial) 
and formation-age (i.e., at the onset of the CV phase) populations that ultimately produced the 
CVs observed in our simulations at the present-day.  We further describe the age-dependences 
found for the properties of our simulated CVs, including effects related to CV destruction and 
escape (from the cluster).

\section{RESULTS AND DISCUSSION}
\label{results}

In the first paper \citep{Belloni_2016a}, 
we were interested predominantly in the properties of 
the PDPs of CVs and the related observational selection effects. In 
this paper, we concentrate on the influence of the cluster dynamics 
in shaping the observed CV PDPs.  We further consider the evolution 
of our CV populations throughout the entire cluster lifetime to 
quantify the predicted age-dependent properties of CVs in GCs.

We begin with a brief summary of the expected present-day CV properties in GCs. 
\citet[][see their table 2]{Belloni_2016a} provide detailed information regarding 
the influence of dynamics in CV formation, for CVs that form during
the cluster evolution and survive to the present-day.  
Their results can be summarized as follows: 

\begin{description}
\item[(i)] all CVs formed in models that adopt a Kroupa IBP have a dynamical 
origin; exchange interactions are the most important channel for CV 
formation, which are responsible for $\sim$ 2/3 of the entire PDP of CVs;
\item[(ii)] for sparse clusters, in particular the S1 and K1 models, 
`dynamical' formation channels occur very rarely;
\item[(iii)] dynamically formed CVs tend to have more massive WDs, due to
exchange interactions that preferentially eject low-mass objects;
\item[(iv)] CVs formed without any influence from the cluster dynamics tend to have 
less massive WDs ($\sim$ 0.3 M$_\odot$); 
\item[(v)] $\sim$ 87 per cent of the CVs formed in our simulations are period 
bouncers, with extremely low-mass donors ($\lesssim$ 0.05 M$_\odot$).  
\end{description}

In the subsequent sections, we elaborate and expand these results.  
We first state the main properties of, and discuss the influence 
of dynamics on, the progenitor populations (Section 
\ref{PP}). Then, we outline the main channels for CV formation 
and compare our results to those of \citet{Ivanova_2006} 
(Section \ref{PP-FAP}). To address the importance of the global cluster 
evolution for CV formation, in Section \ref{GC-EVOL} 
we show that the rate of CV formation is roughly independent
of the cluster configuration throughout its entire lifetime.  
We present the formation-age population properties in Section
\ref{FAP}, and discuss the main implications of these properties 
for observations of CVs in GCs.  This leads to a discussion of the 
influence of dynamics on CV evolution, from CV birth until the 
present-day (Section \ref{FAP-PDP}).  We go on to discuss the 
time-dependent properties of CV populations in GCs, pointing 
out the expected differences between observations of Galactic 
field and cluster CVs (Section \ref{AGE}).  Finally, we discuss 
those CVs formed throughout the cluster evolution, but destroyed 
 or ejected from the cluster before the present-day (Section \ref{DEST})
and CVs formed from binary progenitors that previously escaped 
their host cluster (Section \ref{ESC}).

\subsection{Progenitor Population}
\label{PP}

The progenitor population is identified in CATUABA as follows.  For 
CV progenitors that never undergo exchanges the components of the 
primordial progenitor binary are always the same as the components of the CV, 
so that they are easily identified.  However, the identification is not as 
straightforward when an exchange encounter has occurred.  Here, the 
initial binary properties are obtained by first identifying the properties of the 
binary with the smaller period (if an exchange took place via a binary-binary 
interaction) or simply the properties of the initial binary (if an exchange happened via a 
single-binary interaction).  In the event that both binary components 
of a given CV in the PDP were originally single stars at the time of cluster 
birth, then the CV is excluded from inclusion in the progenitor population, since 
there is no associated initial binary \citep[][see Section 3.1.2 for more details]{Belloni_2016a}.

\begin{figure*}
   \begin{center}
    \includegraphics[width=0.362\linewidth]{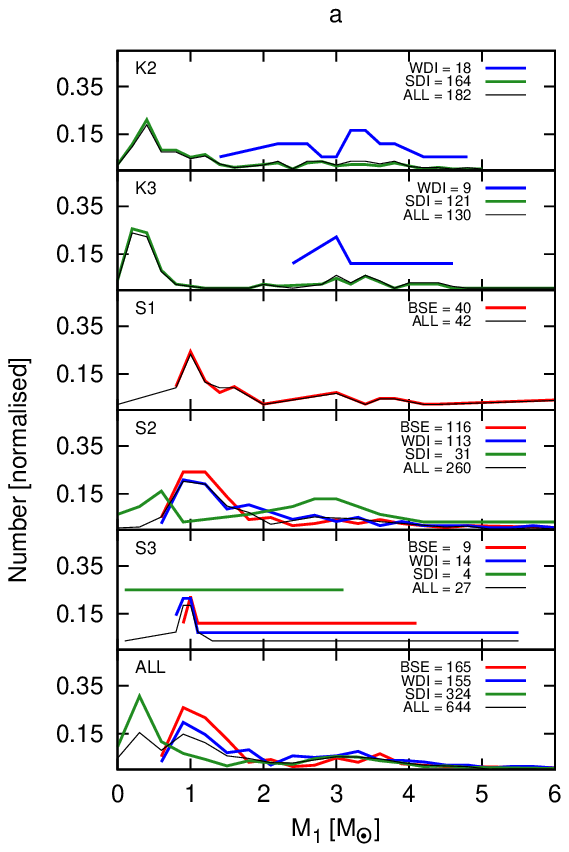} 
    \includegraphics[width=0.315\linewidth]{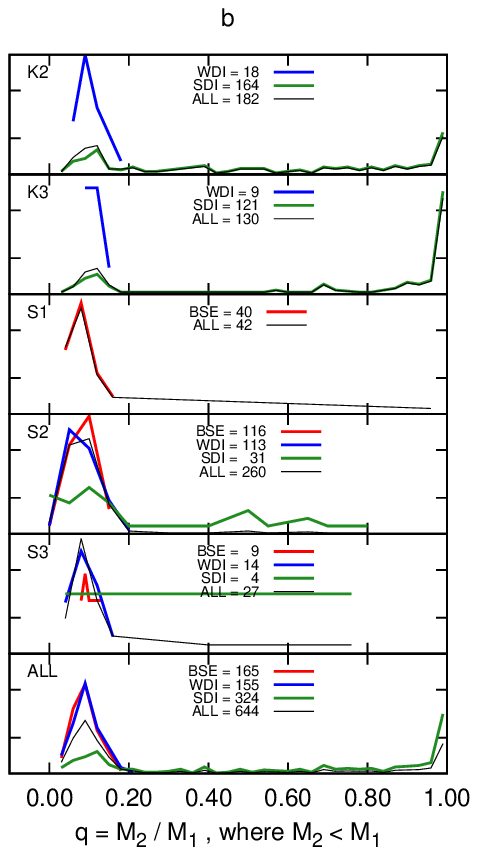} 
    \includegraphics[width=0.315\linewidth]{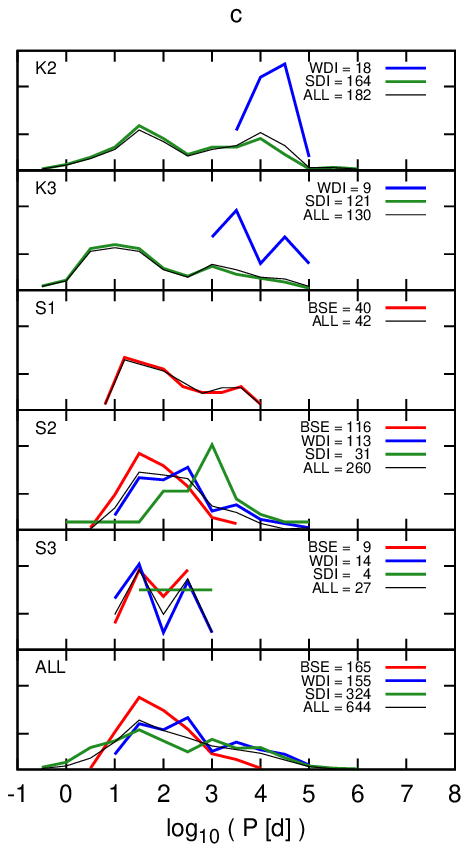} 
    \end{center}
  \caption{Primary mass (left-hand panel), mass ratio (middle panel), and period
(right-hand panel) distributions for progenitor binaries that become CVs and survive to 
the present-day, shown for all models (model K1 is not shown because only three CVs form 
and survive up to 12 Gyr in this model).  
The keys indicate the number of CVs in each model and group.  The 
bottom row presents the progenitor CV populations in all six models 
aggregated into a single sample.  From the bottom row of each graph, 
we see a clear extension in the range of binary parameters relevant to 
CV formation, due to pure binary stellar evolution only (BSE group) being influenced by
dynamical interactions (WDI and SDI groups). 
For more details, see Section \ref{PP}.}
  \label{Fig01}
\end{figure*}

Fig. \ref{Fig01} shows the distributions of the WD progenitor masses, 
mass ratios ($\leq 1$) and periods of the progenitor CV population, 
divided according to the subgroups defined in 
\citet[][see section 3.2]{Belloni_2016a}.

In general, CV progenitors from the BSE group occupy a smaller volume 
in the parameter space defined by the orbital period, eccentricity,
primary mass and mass ratio relative to the WDI and SDI groups.  This is 
because dynamical interactions are able to trigger CV formation in binaries that 
otherwise would never undergo a CV phase, as illustrated in the bottom row of 
each panel in Fig. \ref{Fig01}.

We turn now to a more detailed description of the binary properties characteristic of 
the progenitor populations.  For simplicity, we consider binaries from all six models 
aggregated into a single distribution (bottom row of each panel in Fig. \ref{Fig01}), 
unless otherwise stated.

\subsubsection{Primary Mass}
\label{pp_primary_mass}

Binaries in the BSE and WDI groups are quite similar in terms of their
primary mass (i.e., WD progenitor) distributions (Fig.
\ref{Fig01}a). This indicates that weak dynamical
interactions are not important in changing the mass of the primary.
More precisely, the primary masses (at $t$ = 0) in the BSE
and WDI groups are always more massive than 0.8 M$_\odot$, with a
peak at around 1.0 M$_\odot$ and a long tail towards higher masses
(up to $\sim$ 7.0 M$_\odot$ ). This lower limit for the primary mass
(i.e., WD progenitor at t = 0) indirectly reflects the turn-off mass
evolution, from $t$ = 0 to $t$ = 12 Gyr. As the cluster evolves and
MS stars become compact objects due to stellar evolution, the cluster turn-off
mass shifts toward lower masses. At 12 Gyr (present-day), the
turn-off mass is $\sim$ 0.8 M$_\odot$. This means that MS stars whose
masses are lower than $\sim$ 0.8 M$_\odot$ (at $t = 0$) will not evolve into
WDs within 12 Gyr of stellar evolution.

For binaries in the SDI group, we see primaries with
masses $\lesssim$ 0.8 M$_\odot$, with a peak at $\sim $0.5 M$_\odot$. 
As we have just pointed out above, such low-mass MS stars will not evolve into WDs
within the Hubble time. Thus, those binaries with primary
masses below 0.8 M$_\odot$ (at $t$ = 0) will have at least one component
replaced with a more massive star (either a MS star or a WD)
(see section \ref{fap_exchange}) at a time $t$ > 0. Strictly speaking, for binaries
involved in dynamical interactions that lead
to exchanges, Fig. \ref{Fig01}a does not show the `real' WD progenitor at
$t$ = 0. Instead, Fig. \ref{Fig01}a shows that for this range of primary
masses (at $t$ = 0), the binaries must be involved in at least one
exchange interaction (at $t$ > 0) in order to form a `real' CV
progenitor binary. Said another way, Fig. \ref{Fig01}a shows that most binaries
in the SDI group cannot form CVs as they are at $t$ = 0. Thus,
exchange interactions are needed to convert
such binaries (at $t$ = 0) into `real' CV progenitors (at $t$ > 0).
This is especially true for models K2 and K3, in
which exchange interactions are the principal CV formation channel.
We discuss the role of exchanges in CV formation in more
detail in Section \ref{fap_exchange}.

\subsubsection{Secondary Mass}
\label{pp_secondary_mass}

Analogous behaviour is observed between the BSE and WDI groups 
for the secondary mass distribution (not shown in Fig. \ref{Fig01}).  
The reasons for this are similar as for the primary mass: only
strong dynamical interactions can change the component masses.

The masses are, usually, 0.08 < $M_{\rm 2}$ < 1.0 ${\rm M_\odot}$ in the WDI 
group, with only two cases in which the secondary mass is greater than 
1.0 ${\rm M_\odot}$. One of these binaries forms an unstable CV, that later becomes 
stable due to mass loss. 

\subsubsection{Mass ratio}
\label{pp_mass_ratio}

The mass ratio distribution (Fig. \ref{Fig01}b) shows 
a clear separation between those CVs formed with a strong dynamical 
influence and the rest.  For binaries in the BSE and WDI groups, the initial
mass ratio is always $\lesssim$ 0.2.  Otherwise, the outcome of the
CEP is an unstable CV with a high probability of merging 
after a few Myr due to an enhanced mass transfer rate. However,
in some extremely rare cases, a few CVs were born unstable and later 
became stable due to mass loss.  In these rare cases, the mass ratio after 
the CEP is $\sim$ 1. Of all six models considered, only one such case emerged, 
in model S2.  Typically, unstable systems do not survive the mass transfer process,
as we will see in more detail in Section \ref{unstable_CVs}.

Now, for the CVs in the SDI group, the mass ratio distribution covers the full range, 
with a slight preference for q < 0.15 (in models S2, K2, and K3),
and q $>$ 0.95 (in models K2 and K3). This extensive range is due to the 
fact that binaries can undergo exchanges, which can significantly change their
properties. This is clear from Fig. \ref{Fig01}b.  Additionally,
models K2 and K3 show a peak around q $\sim$ 1. This
is because the Kroupa models have an initial mass ratio 
distribution with its peak $\sim$ 1.

\subsubsection{Period}
\label{pp_period}

The upper limit of the period distribution (Fig. \ref{Fig01}c) 
for the BSE group is always less than $\sim 10^4$ days. 
This is because secondaries in post-common-envelope binaries (PCEBs)
descended from longer period binaries will not fill their 
Roche-lobes within the Hubble time.

The WDI group contains binaries that suffered only weak 
dynamical interactions.  For these binaries, the initial period can
be longer than 10$^4$ days, if the eccentricity is increased via 
multiple dynamical interactions.  Specifically, the upper period limit can 
reach $\sim 10^5$ days, particularly for models K1 and K2.
Such models mostly contain long-period binaries in the WDI group, 
which indicates that the cumulative effect of weak dynamical 
interactions is to significantly change the eccentricity, 
pushing wide binaries to become CVs.  
On the other hand, models S2 and S3 mostly have short-period 
binaries, which suggests that the cumulative effect of weak
dynamical interactions is insignificant. However,
in a few rare cases, this cumulative effect will compound to become 
significant might be strong (note that a few binaries have periods longer 
than $\sim 10^4$ days in model S2).

The SDI group shows no evidence for a restricted interval in the period distribution, 
since the binary periods are changed drastically due to strong interactions.  
We do, however, see an extension of the period interval 
(compared to the BSE and WDI groups) 
in the direction of both shorter and longer periods.

\subsubsection{The Progenitor Population}
\label{pp_discussion}

To sum up, we see a clear correlation between an expansion in the
range in CV progenitor parameter space 
and the strength of dynamical interactions.  In other words, dynamical
interactions can bring binaries to the required range in parameter
space for CV formation to occur.

\citet[][]{Ivanova_2006} found similar results, although
in their Fig. 1, they considered only binaries with circular orbits.  This 
originated from a very restricted IBP, that was evolved without
the influence of dynamics.  Consequently, we are only able to compare 
their findings to our results for the BSE group.

\subsection{From progenitor up to formation}
\label{PP-FAP}

After describing the properties of the progenitor binary population, 
we now move on to investigating the main formation scenarios leading to 
CV formation. 

\subsubsection{Main formation channels}
\label{fap_formation_channels}

Table \ref{Tab1} lists the main formation channels for CVs in our simulations.  
For comparison, we also included the corresponding 
formation channels described in \citet{Ivanova_2006}. Briefly, we have 
CV formation through CEP with/without weak/strong dynamical interactions,
CV formation through exchanges with/without CEP, and CV formation
through mergers with/without CEP.  
We are able to reproduce with MOCCA only some 
of the channels in \citet{Ivanova_2006}, since in MOCCA tidal capture 
(including collisions between red giants (RGs) and MS stars that lead to the 
formation of WD-MS binaries) is not modeled. 

We define our formation channels as follows: 

\begin{description}
\item[(1)] CEP without any dynamical interactions, either before or after the CEP;
\item[(2)] CEP with weak/strong dynamical interactions, either before or after the CEP;
\item[(3)] exchange without any dynamical interactions post-exchange; 
\item[(4)] exchange with dynamical interactions post-exchange;  
\item[(5)] exchange followed by a CEP;   
\item[(6)] merger without any dynamical interactions post-merger; 
\item[(7)] merger with dynamical interactions post-merger;  
\item[(8)] merger followed by a CEP post-merger. 
\end{description}

\begin{table}
\centering
\caption{Number of CVs present in the clusters at 12 Gyr
separated according to their main formation channel. The channels in 
\citet[][see their Fig. 7]{Ivanova_2006} correspond to the second row.
The numbers given in the fourth row indicate the CV formation channel identified
in this work. See Section \ref{fap_formation_channels} for more details.}
\label{Tab1}
\begin{adjustbox}{max width=\linewidth}
\noindent
\noindent
\begin{tabular}{l|c|c|c|c|c|c|c|c|c}
\hline\hline
 & \multicolumn{9}{c}{Formation Channel as defined by \citet{Ivanova_2006}} \\
\hline\hline
 & 1a,c & 1b & 2a & 2b & 2c & 3a & 3b & 3c & \\ 
\hline\hline
 & \multicolumn{9}{c}{Formation Channel as defined in this work} \\
\hline\hline
Model & 1 & 2 & 3 & 4 & 5 & 6 & 7 & 8 & Total\\ 
\hline\hline
K1 &          0 &      1 &    0 &    0 &    2 &    0 &     0 &     0 &    3 \\ \hline
K2 &          0 &     61 &   71 &   13 &   22 &    2 &     1 &    12 &  182 \\ \hline
K3 &          0 &     31 &   57 &    4 &   10 &    7 &     0 &    21 &  130 \\ \hline
S1 &         40 &      1 &    0 &    0 &    1 &    0 &     0 &     0 &   42 \\ \hline
S2 &        122 &    124 &    5 &    1 &    5 &    0 &     0 &     3 &  260 \\ \hline
S3 &          9 &     16 &    0 &    0 &    2 &    0 &     0 &     0 &   27 \\ \hline\hline 
Total &     171 &    234 &  133 &   18 &   42 &    9 &     1 &    36 &  644 \\ \hline 
Fraction & 0.27 &   0.36 & 0.20 & 0.03 & 0.07 & 0.01 & <0.01 &  0.06 & 1.00 \\ \hline 
\hline 
\end{tabular}
\end{adjustbox}
\end{table}

\subsubsection{Comparisons with \citet{Ivanova_2006}}
\label{fap_formation_channels_comp}

From the numbers in Table \ref{Tab1}, we see that our results 
agree roughly with those found by \citet{Ivanova_2006}. 
Here, we discuss only a restricted 
number of models, by considering only those of our models 
with initial conditions that match those in \citet{Ivanova_2006}.

The models in \citet{Ivanova_2006} with low densities yield results very close 
to our models K1 and S1.  \citet{Ivanova_2006} showed that most CVs in such low density clusters
form through a CEP without any influence from dynamics. We also
found this for our model S1.  Model K1, on the other hand, is more difficult to compare.  
This is because it is a Kroupa model and, for the binary stellar
evolution parameters adopted in this investigation, no 
CVs formed in K1 without any influence from dynamics \citep{Belloni_2016a}.

Additionally, our `aggregated/average' cluster is similar
to their Standard model (compare last row of Table \ref{Tab1} and
the first row in their Table 2). We also found that only $\sim$ 27 per cent
of CVs are formed purely through a CEP without any contribution from dynamical interactions.
The rest of the CVs were involved in some kind of interaction(s).  
The channels connected with exchanges and mergers are also quite similar
in both models (our `aggregated' and their Standard). 
However, there are two important differences that we must mention.

First, we cannot compare the influence of tidal capture and the physical
collision of an MS star with an RG, because such a feature is not implemented in 
MOCCA.  Secondly, we have an average enhancement with respect to formation
channel 2, relative to their results. This could be associated with the fact
that they only analyzed CV formation in GC cores, and that
MOCCA's cluster models evolve in time, while theirs do not.

The following issues should be kept in mind for meaningful 
interpretations of our results.
\begin{description}
\item[(i)] We have a restricted number of models/initial conditions. Our
models do not match those in their study, with a few exceptions.
\item[(ii)] We have evolving clusters. Their models remain frozen in time.  Consequently, 
they may lack some effects related to the cluster dynamical evolution.
\item[(iii)] We have CV formation throughout the cluster. They considered 
only CV formation inside the core.
\item[(iv)] We do not have information about the influence of tidal capture.  \citet{Ivanova_2006}, 
however, have ruled out this mechanism as an efficient way
to produce CVs in GCs.
\item[(v)] We use the CEP efficiency given by \citet{Hurley_2002},
i.e. $\alpha$ = 3.0, which might be inappropriate \citep{Zorotovic_2010}. They
use a CEP parametrization, such as $\alpha \lambda = $ 1.0, which
also does not seem to be appropriate \citep[e.g.][]{Camacho_2014,Zorotovic_2010}.  
This is because the common-envelope binding energy $\lambda$ has different values
for different giants.
\end{description}

All of these effects should be kept in mind when performing comparisons between our study 
and that of \citet{Ivanova_2006}.  In general, as already pointed
out, a general and rough overall agreement is found between these two investigations.  
A more thorough comparison will be performed in a forthcoming paper, that 
will include a larger sample of MOCCA simulations.

\subsubsection{The role of exchange interactions}
\label{fap_exchange}

Dynamics exchanges more massive CVs into the
PDP. This was already noted by \citet[e.g.][]{Ivanova_2006}.  
Typically, a binary composed of low-mass MS stars will have one
of them replaced by either a more massive MS star or a more massive WD. In 
the former case, the binary will eventually undergo a 
CEP, producing products with higher masses (at least for the WD mass). This happens in 
roughly 20 per cent of such exchange interactions.
In 70 per cent of CVs formed due to exchanges, a low-mass MS star is replaced by a more massive WD
and there is no CEP.  Mass transfer starts after circularization and/or dynamical interactions
and/or angular momentum loss.  In the remaining 10 per cent, the pre-exchange binaries 
were composed of an RG and an MS star. In these cases, the MS star is replaced by another MS
star with similar mass, followed by a CEP.

\citet{Ivanova_2006} found similar results. They found that $\sim$ 20 per cent of
the CVs formed post-exchange underwent a CEP (the last exchange led to an 
MS-MS binary), $\sim$ 80 per cent of CVs formed post-exchange did not
pass through a CEP (the last exchange led to an MS-WD binary).

\subsubsection{Acceleration and Retardation of CV formation}
\label{fap_acceleration}

An additional effect associated with dynamical interactions is that they 
can either retard or accelerate CV formation. \citet{Shara_2006} 
noticed evidence for the cluster environment accelerating CV formation
(e.g., their CV6).  
In this work, we confirm this possibility, but also the inverse, i.e. CV formation that 
is retarded due to dynamical interactions. We have
only one case in which there is a strong dynamical interaction after CV formation,
which led to an acceleration of the CV evolution.
This case will be discussed in Section \ref{cv_evolution_with}.

For accelerated CV formation, 
the CV progenitor interacts with single stars (or, more rarely,
with softer binaries), producing a harder (i.e., smaller period) 
and/or a more eccentric binary. Both effects lead to a smaller
pericentre distance.  This causes the dynamically unstable 
mass transfer from the giant to take place earlier, 
when the giant is not so evolved (i.e., it has a small radius). 
This precedes a CEP, and in 
turn CV formation. Such CVs correspond to $\sim$ 30 per cent of all 
CVs that had their formation times altered by dynamical interactions.
Additionally, these CVs have smaller WD masses (since the increase of the 
giant core mass stops earlier) than their field-like counter parts.

For retarded CV formation, the CV progenitor interacts with harder binaries, causing it 
to become softened (i.e., migrating to longer periods). This effect leads to a greater
pericentre distance, which makes dynamically unstable mass transfer from the giant
take place later, when the giant is more evolved (i.e., it has a large radius). 
Thus, the CV progenitor starts the CEP later, which produces a more massive WD 
since the giant core has more
time to evolve and increase its mass. Retardation takes place more
frequently than acceleration, corresponding to $\sim$ 70 per cent of all cases.

To sum up, the dynamical environment can either accelerate or retard CV formation
due to dynamical interactions. 
Although \citet{Shara_2006} performed a systematic study of CV formation with NBODY6,
their investigation suffered from small-number statistics (i.e., only two models).  
Nevertheless, the wealth of detail provided for the few CVs formed in their simulations 
makes it possible to compare their results to ours.

\subsection{The cluster evolution and the CV formation rate}
\label{GC-EVOL}

\begin{figure*}
   \begin{center}
    \includegraphics[width=0.48\linewidth]{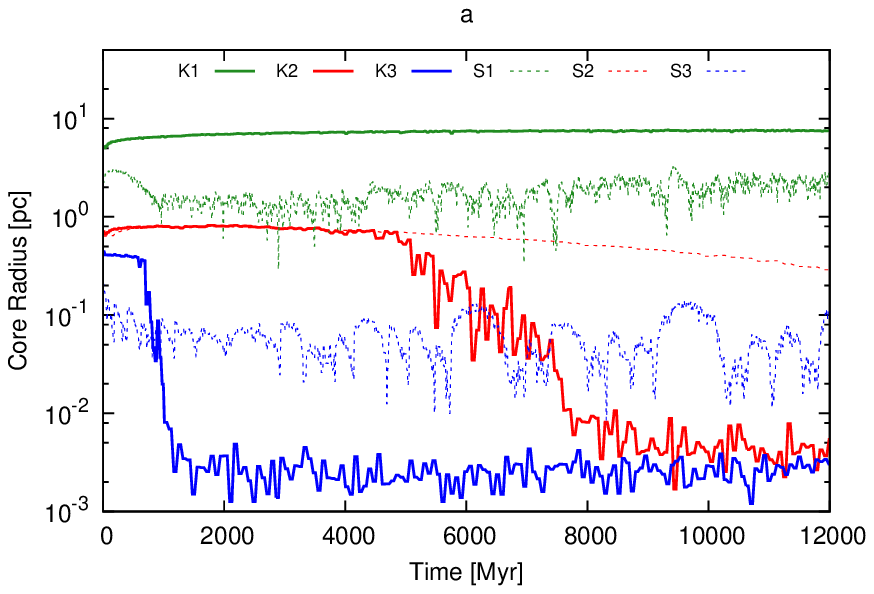} 
    \includegraphics[width=0.48\linewidth]{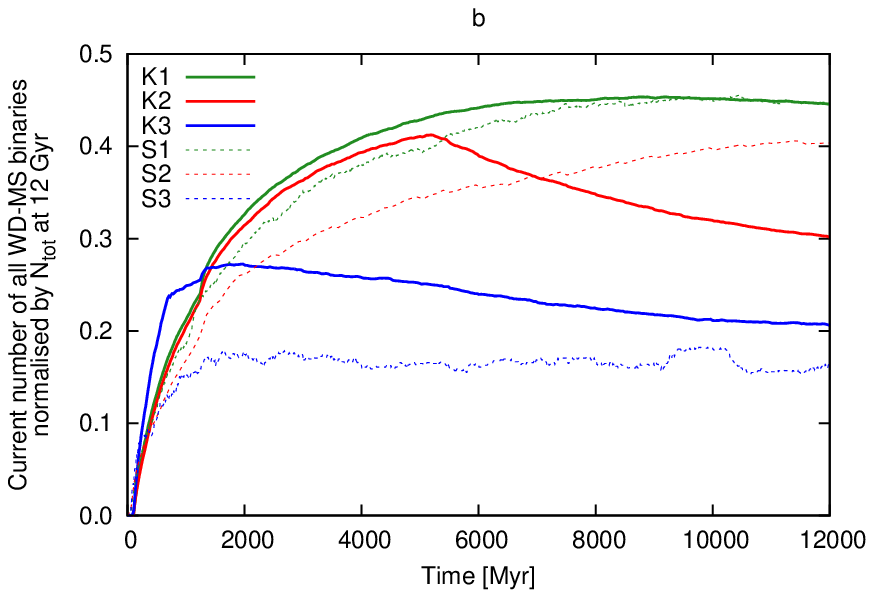} 
    \includegraphics[width=0.48\linewidth]{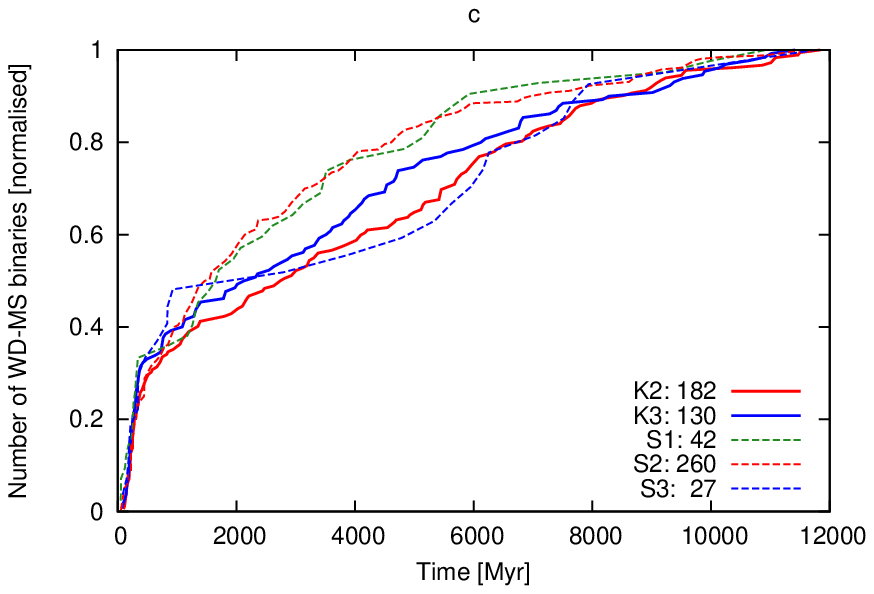} 
    \includegraphics[width=0.48\linewidth]{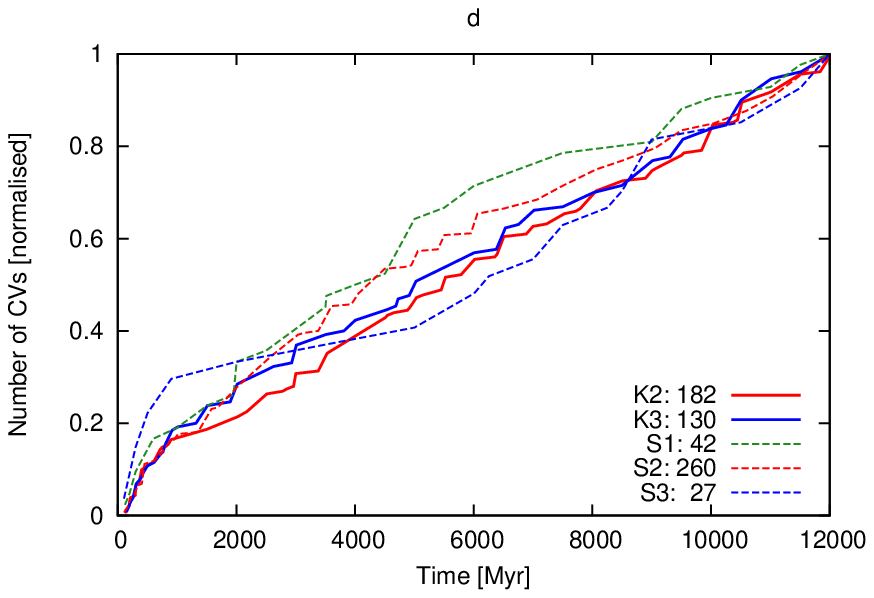} 
    \end{center}
  \caption{{\it Top left}: core radius ($r_{\rm c}$) evolution, computed according
to \citet{Casertano_1985}. Note that, at 12 Gyr, there are 4 
post-core collapse clusters (K2, K3, S1 and S3); two pre-collapse 
clusters (K1 and S2); and two clusters with intermediate-mass 
black holes (K2 and K3). {\it Top right}: number of detached WD-MS binaries
in the cluster at particular times, which corresponds to the total number
of WD-MS binaries formed minus the total number of WD-MS binaries destroyed
up to particular times, normalized with respect to the total number of WD-MS
binaries formed up to 12 Gyr. {\it Bottom left}: cumulative number of WD-MS binaries
that will at some point form the present-day CV population, normalized by the number of present-day
CVs. {\it Bottom right}: cumulative number of
present-day CVs, normalized by the number of present-day
CVs. Notice that model K1 is not shown because it has only 3 present-day CVs.
For details, see Section \ref{GC-EVOL}.}
  \label{Fig03}
\end{figure*}

We show in this section that the CV formation rate can be divided 
into two distinct regimes: a burst in the beginning ($\lesssim$ 1 Gyr) 
followed by a nearly constant formation rate after $\sim$ 1 Gyr, 
regardless of either the initial cluster properties (including the initial
binary fraction, initial binary population, and initial central density), 
the cluster evolution or the strength of dynamical interactions during the 
formation process.

\begin{figure*}
   \begin{center}
    \includegraphics[width=0.48\linewidth]{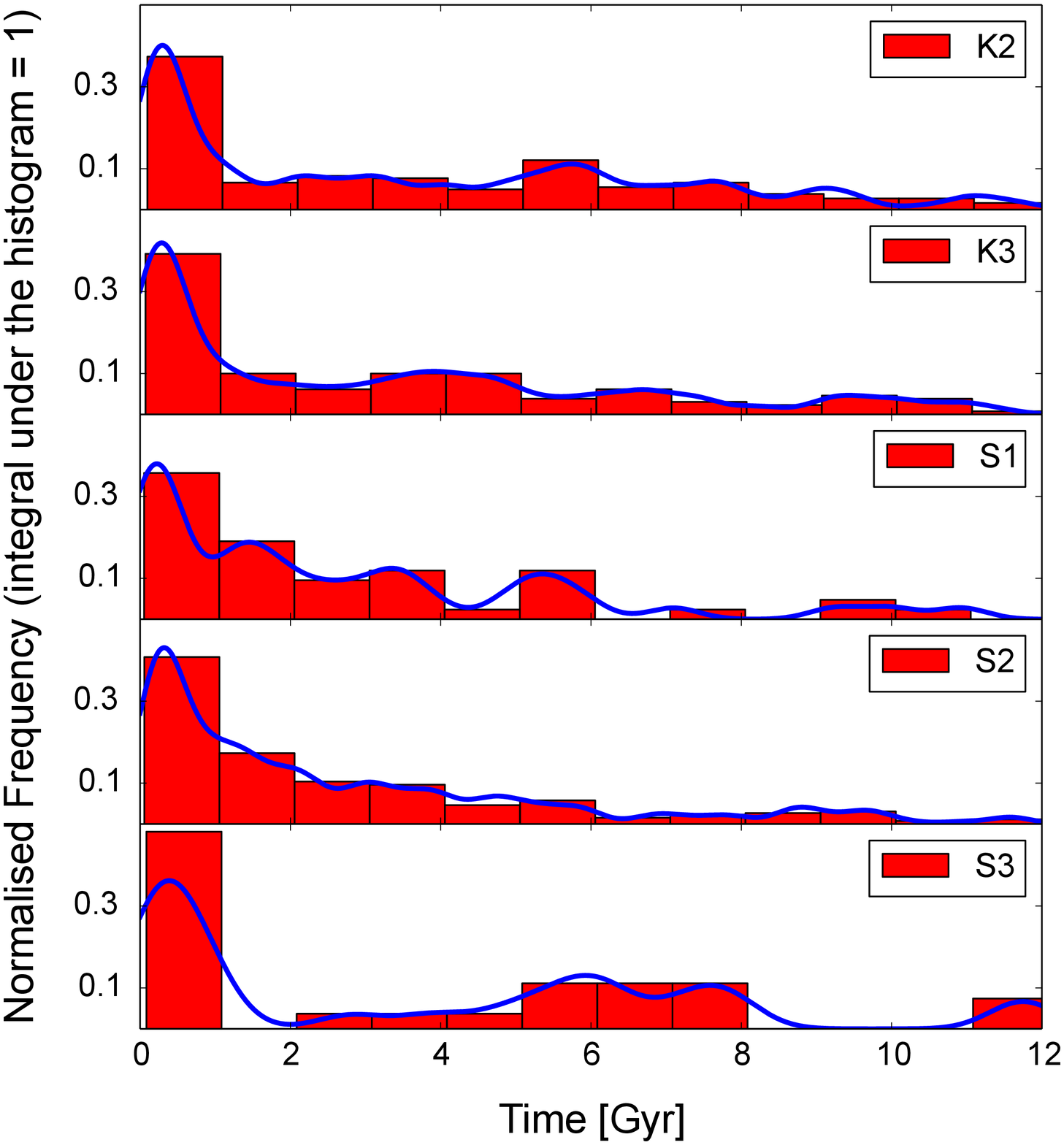}
    \includegraphics[width=0.48\linewidth]{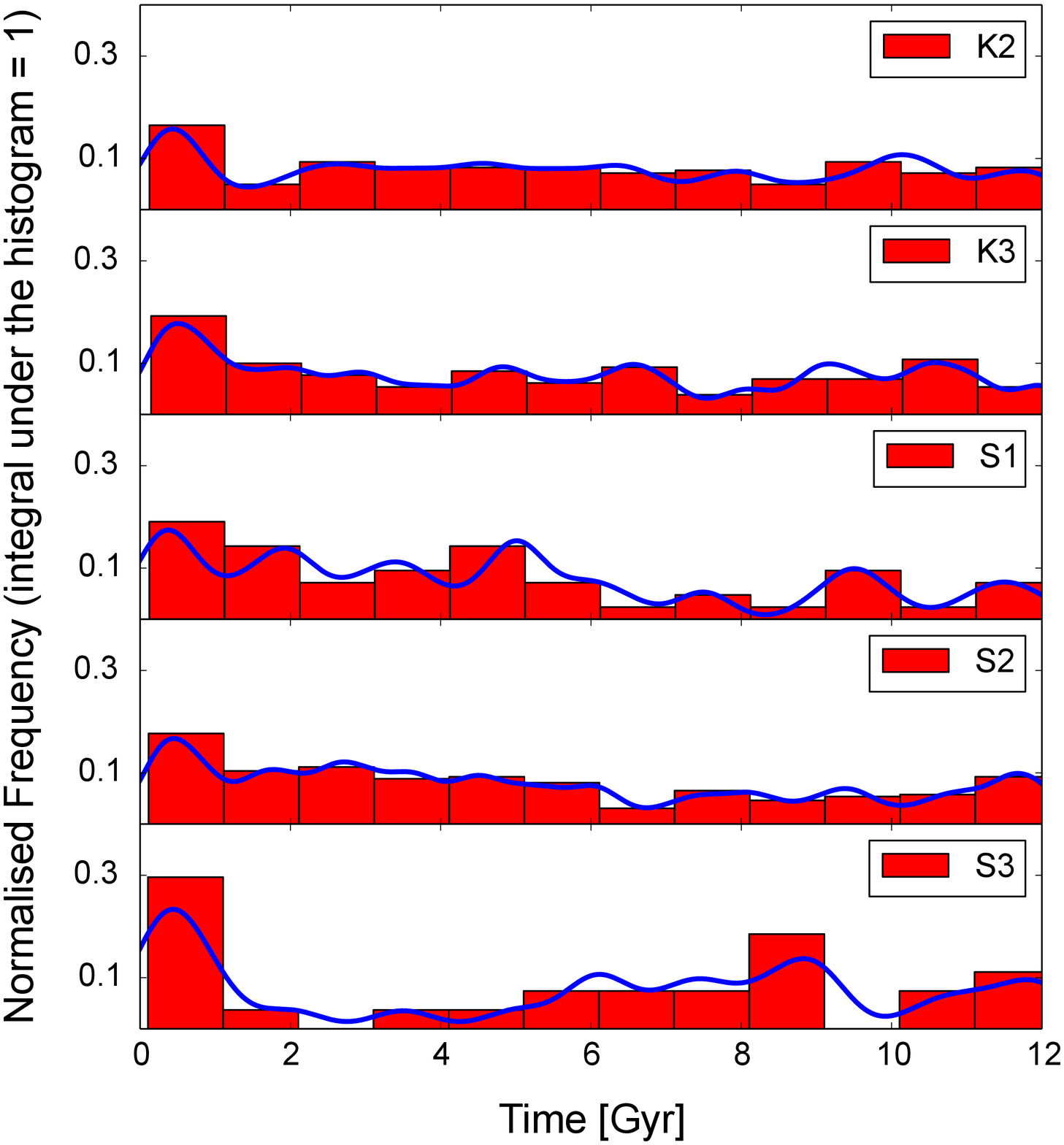}
    \end{center}
  \caption{{\it Left}: formation rate ($\Delta N/\Delta t$) of 
WD-MS binaries that will later become the present-day CV populations in five models.
{\it Right}: formation rates ($\Delta N/\Delta t$) of the present-day CVs in the five models. 
In both panels, the line corresponds to the probability density 
function estimated by the kernel density estimation method using
Gaussian kernels with a bandwidth of $\sim$ 0.1.
Notice that model K1 is not shown because it has only 3 present-day CVs.
Note also that the WD-MS binary formation rate and the CV formation rate show an initial
burst, followed by smooth monotonic decrease.  However, 
the CV formation rate is flatter (nearly constant), due to a 
time-delay between WD-MS binary formation and CV formation.}
  \label{Fig03.1}
\end{figure*}

\begin{figure}
   \begin{center}
    \includegraphics[width=220px]{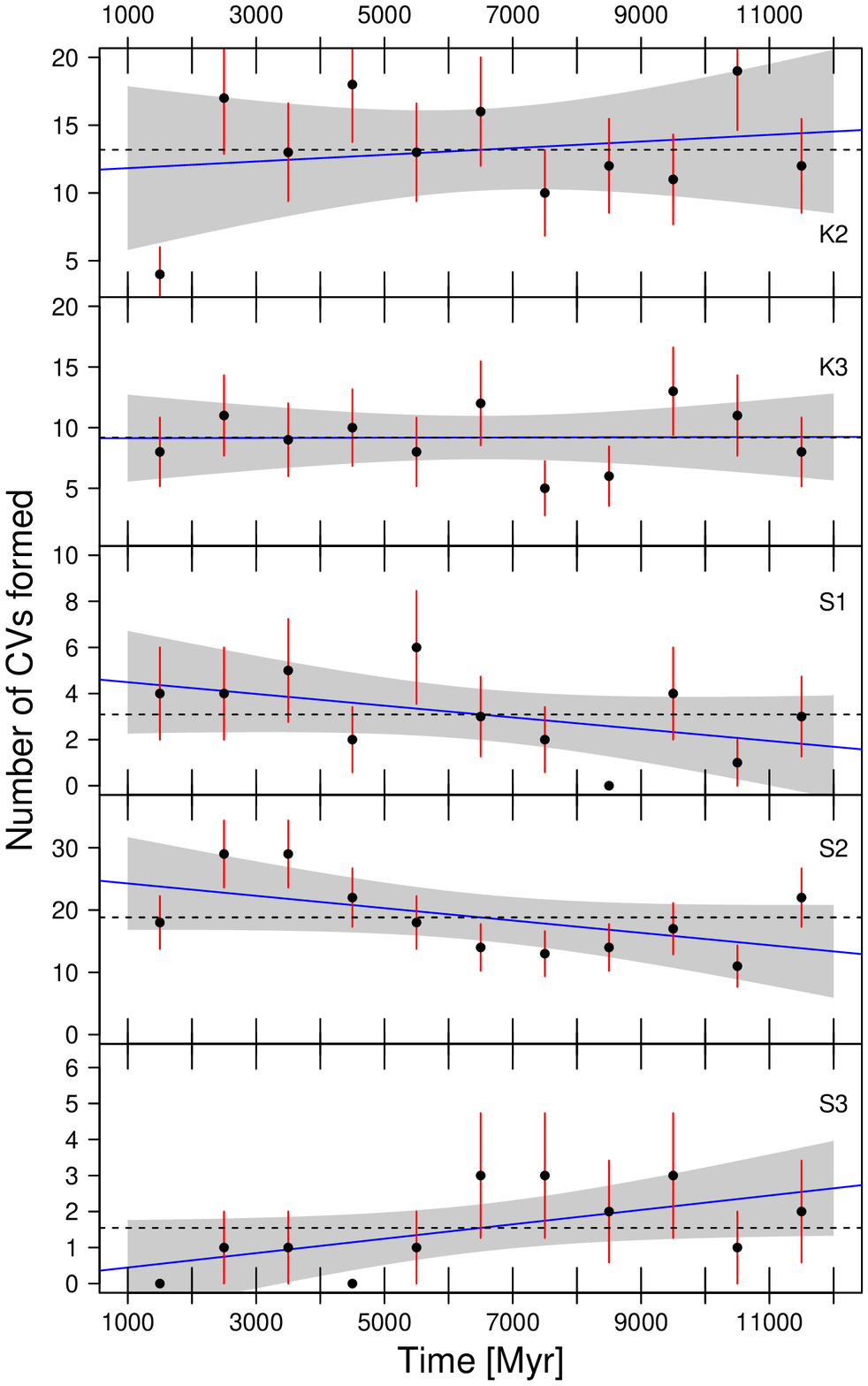}
    \end{center}
  \caption{CV formation rate. Points represent the number of CVs formed after 1 Gyr, 
in intervals of 1 Gyr. The error bar represents Poisson errors ($\sqrt{n}$), 
and the blue line is the linear regression model ($N_{\rm for}=a + bt$, where
$N_{\rm for}$ and $t$ are the number of CVs formed and the time, respectively).
The gray area is the band of 95\% confidence interval around the model. 
The dashed horizontal line is the average rate, assuming a constant
CV formation rate (i.e. $b=0$). Note that all dashed lines are inside
the gray bands, which indicates that all models are compatible with 
a constant formation rate. However, models S2 and S3 are in the extreme of 
acceptability, since their parameters $b$ are barely significant 
($p$-values 0.08 and 0.06, respectively, but still not higher than the 
classical 0.05 for statistical significance).}
  \label{Fig03.2}
\end{figure}

\subsubsection{Cluster Evolution}
\label{GC-EVOL-cluster}

In this section, we discuss the time evolution of the cluster properties 
in order to quantify the impact on CV formation.

Fig. \ref{Fig03}a shows the behaviour of the 
core radius, calculated according to \citet{Casertano_1985}. 
Recall that the six models can be roughly described as: two sparse (K1 and S1), 
two dense (K2 and S2) and two very dense (K3 and S3) models; 
four post-core collapse (S1, S3, K2 and K3) models, and two with 
intermediate-mass black holes (K2 and K3).

For model K1, the cluster is expanding
and has a long relaxation time-scale. This is due to the 
cluster's large size, causing it to fill its tidal radius with an extremely low 
central density.

On the other hand, model S1 is comparably large, with a low density, and it 
experienced an episode of core-collapse around 1 Gyr.  This model is particularly
interesting because pre-core-collapse, it managed to form a black 
hole subsystem in the core. \citet{Breen_2013} explain the evolution of this kind 
of core collapse; the energy generated 
in the core is regulated by two-body relaxation throughout the entire system.  In 
clusters with a centrally concentrated black 
hole subsystem (without a massive central black hole), the energy is 
generated by three-body encounters in the core.  
As the black holes escape from the cluster, it undergoes another 
episode of core-collapse.  Once the balanced evolution is restored, 
the core of low-mass stars needs to be compact to allow for 
the formation and interaction of binaries composed of less 
massive stars.  It is worth mentioning that the lifetime of the
black hole subsystem is determined by the relaxation time-scale 
of the cluster, not by the dynamical time-scale of the black hole subsystem,
which is much shorter. For more details on this particular
scenario, see \citet{Wang_2016}.

We note that for models with intermediate-mass black holes (K2 and K3),
the prescription given by \citet{Casertano_1985} for the core radius begins to break down.  
This is because the mass of the core is significantly affected 
by the intermediate-mass black hole, negatively affecting the determination of the 
core radius.  Again, the black holes provide 
a source of energy for clusters K2 and K3. Additionally, the intermediate-mass
black holes in these models are formed in the way explained by \citet{Giersz_2015}.
The intermediate-mass black hole in model K2 forms quite early ($\lesssim$ 1 Gyr), whereas 
the one in model K3 forms at $\sim$ 5 Gyr.

Model S2 is approaching core-collapse at the present-day.  The collapse will occur 
at 13 Gyr, such that the core density in this cluster is currently increasing, and quickly.  

Finally, model S3 has initially only 300k stars, although it is extremely
dense. This makes the relaxation time short, and core-collapse
takes place very early on (several Myr after the start of the simulation).

\subsubsection{Reservoir of WD-MS binaries}
\label{GC-EVOL-wdms}

Before discussing the CV formation rate, 
we stress that WD-MS binary formation and CV formation are
two distinct processes.  {\it WD-MS binary formation} involves the formation
of a detached binary composed of a WD and an MS star (due to a CEP, exchange, merger, etc.). 
{\it CV formation} involves the formation of a semi-detached WD-MS binary,
i.e. an interacting WD-MS binary.  Both formation channels can occur at any time, a priori. For 
instance, a massive PCEB (formed in the first few Myr of the cluster
evolution) can become a CV at any time, depending on its orbital period and on its
MS mass, just after the CEP. 
If the period is small and the MS mass is great, then CV formation takes place fast. 
On the other hand, if the period is long and the MS mass is small, then the PCEB
might have to wait a few Gyr (in order for its orbit to shrink due to angular momentum
loss) until it initiates the CV phase.

We now move on to a brief analysis of the number of WD-MS binaries in 
clusters as a whole, i.e. let us first concentrate on all WD-MS binaries present
in our simulated clusters throughout their evolution.

With respect to the formation of WD-MS binaries in the six models, two
features are worth highlighting: an initial burst followed by a smoothly
decreasing formation rate.  This is due to the MS lifetime. Since the main-sequence lifetime  
is inversely related to mass, we see a 
very fast formation rate of WDs early in the cluster evolution (due to the
fast evolution of massive MS stars), followed by a smoothly decreasing
formation rate (due to the evolution of MS stars whose masses are below
$\sim$ 2.0 M$_\odot$ and spend more time on the MS).

Fig. \ref{Fig03}b shows the number of
WD-MS binaries in the clusters normalized by the total (cumulative)
number of WD-MS binaries formed up to 12 Gyr.  We find that models 
K1, S1, and S2 show a similar behaviour as described above, since the
number of WD-MS binaries is continuously increasing.  Hence, in these three 
models, this is reflected in the 
number of WD-MS binaries, which implies in turn that most WD-MS binaries formed
in clusters are not destroyed (not strongly affected by dynamical interactions).

On the other hand, for models K2 and K3, we see that, at a particular time,
the destruction of WD-MS binaries becomes more prominent.  This is evidenced by the fact that the
number of WD-MS binaries, from this point on, starts to decrease.  
Interestingly, the times at which the destruction process begins to dominate are 
precisely the times corresponding to the formation of their intermediate-mass black holes, in both 
models.  This is consistent with the work of \citet{Leigh_2014} and \citet{Giersz_2015}.  The 
intermediate-mass black holes form at $\sim$ 1 and $\sim$ 5 Gyr, for models K2 and K3, 
respectively.

Another interesting feature comes from comparing the K models collectively to 
the S models. Notice that the K curves in Fig. \ref{Fig03}b are
always above the S curves, when the WD-MS binary formation rate surpasses
the destruction rate, in all models. This is a consequence of the period
distribution of the initial binaries.  In the S models, the relative
number of hard binaries is greater, in comparison with the K models
(that have predominantly wide binaries). This makes mergers 
during or after a CEP more frequent for PCEBs in the S models.  
This way, even when the production rate dominates, many binaries
merge on their way to becoming WD-MS binaries.

Finally, model S3 shows a weak destruction rate of WD-MS binaries up
until $\sim$ 1 Gyr. After that, we see clearly that the number
of WD-MS binaries in the cluster remains roughly constant, which
means that the WD-MS binary formation rate balances the destruction rate.

All of the aforementioned processes reflect on the rate at which 
WD-MS binaries become the present-day CV population. Additionally, 
we see clearly from Fig. \ref{Fig03}b
that dynamics brings about a different evolution in the 
number of WD-MS binaries in each type of cluster, which 
provides a clue to the dynamical age of the cluster.

\subsubsection{Formation rate of WD-MS binaries that are CV progenitors}
\label{GC-EVOL-wdms}

Fig. \ref{Fig03}c shows the cumulative number of
WD-MS binaries that will later become the present-day CV population.  
We can separate the WD-MS binary 
formation rate into two distinct regimes: an initial burst that
lasts up to $\sim$ 1 Gyr, followed by a roughly smoothly decreasing 
formation rate.  This is analogous to the formation rate of 
the WD-MS binary populations in our simulated clusters.  
This is apparent from the left-hand panel of Fig. \ref{Fig03.1}, which
shows this formation rate. 

We also find that models S1 and S2 show a more prominent decrease 
in the CV progenitor formation rate relative to models K1 and K2
(compare first/second and third/fourth rows of Fig. \ref{Fig03.1}).  
This is associated with the cluster density, and it reflects the intimate balance 
between formation and destruction.  Models S1 and S2 are less dense than models
K2 and K3.  This makes the destruction of potential CV progenitors less
frequent in models S1 and S2.  These two different behaviours (one for models S1 and
S2, and the other for models K2 and K3) are intrinsically associated
with the influence of dynamics.  In models S1 and S2, the production of WD-MS binaries 
is driven by stellar evolution processes.  On the other hand, models K2 and K3 
are much denser and are post-core collapse clusters, which increases the dynamical 
production rate of WD-MS binaries. 
Nevertheless, we see a general pattern associated with an initial 
burst in the formation rate, followed by a roughly smoothly decreasing rate. 
The intensity of such a decrease depends on the main CV formation channel.
For CVs associated with CEP (with/without weak dynamical interactions), like those
predominantly formed in models S1 and S2, we note a 
more rapid decrease in the formation rate in the first Gyr. On the other
hand, for dynamically formed CVs (models K2 and K3), we see a less rapid 
decrease (compare KDE models in Fig. \ref{Fig03.1}). Then, dynamics tend
to `flatten' the WD-MS formation rate.

Finally, model S3 shows a combination of the above two features,
since this model is extremely dense.  This leads to the presence of 
a significant population of dynamically-formed
CVs.  Meanwhile, CVs also form purely through stellar and binary evolution. 
We see the initial burst of WD-MS binary formation, followed by a peculiar 
time-dependence for the formation rate (see the bottom row of Fig. \ref{Fig03.1}).
Unfortunately, we are not able to say more about this model, since
its number of CVs is very small.

\subsubsection{CV formation rate}
\label{GC-EVOL-cv}

Having explained the WD-MS binary formation rate, we now turn 
to the CV formation rate. 
Once a WD-MS binary forms, there is a time-delay before 
CV formation.  The duration of this delay depends on the period and MS mass of the
WD-MS binary.

Comparing Figs. \ref{Fig03}c and \ref{Fig03}d, we see
that the cumulative numbers of WD-MS binaries and CVs are different.
This difference is due to the above-mentioned time-delay
between WD-MS binary and CV formation. 

In any event, from Fig. \ref{Fig03}d and the right-hand panel of 
Fig. \ref{Fig03.1} (which shows the CV formation rate), we see again an initial burst.
However, for the CVs, a nearly constant formation rate follows the
initial burst (different from the WD-MS binary formation rate, that shows a roughly 
smoothly decreasing formation rate following the initial burst).  
This was already noticed by \citet{Ivanova_2006}, 
who concluded that the cluster evolution is not important
with respect to the {\it relative number of CVs appearing} in GCs.

In order to test this claim, we applied a 
two-sample Kolmogorov-Smirnov test to 
pairwise combinations of all five models.
Since the alternative hypothesis of the test is that 
{\it the paired samples do not stem from the same theoretical 
CV formation rate}, a null result supports the conclusion that 
the CV formation rate is model-independent, in the sense that 
the CV populations form in a similar way irrespective of the 
cluster properties.

None of these tests support the rejection of the null hypothesis 
with 99\% confidence, which suggests that there is no strong 
evidence that the CV formation rate differs from model to model.

However, these results are sensitive to small number statistics.  
In fact, when models K2 and S2 are tested, we can reject the null 
hypothesis that they are similar with a confidence level greater 
than 97\% ($p$-value = 0.02).  For these two models, the KS test 
suggests that we have enough evidence to reject the null hypothesis.  
This is important, since these two models have the largest numbers of CVs. 
It is likely that the uniqueness of the CV formation rate amongst 
different GC models can only be correctly analyzed with larger 
simulations yielding more CVs.

In order to test whether the CV formation rate is indeed nearly constant
after $\sim$ 1 Gyr, we applied a one-sample Kolmogorov-Smirnov
test for uniformity to the same five models.
Only model S2 presents a small $p$-value (1.224$\times 10^{-4}$).
This is because more CVs form in model S2, and it ends up being 
easier to identify deviations in this model, i.e. it is easier 
to notice that model S2 does not follow the other models. 
The other models do seem different, but the 
small number statistics make them technically indistinguishable.

Proceeding further, we performed a linear regression for 
the number of CVs formed after 1 Gyr, in each interval of 1 Gyr,
using the absolute numbers in Fig. \ref{Fig03.1}. If the CV formation rate 
is constant, then a fit of the form $N_{\rm for} = a + bt$, where
$N_{\rm for}$ is the number of CVs formed and $t$ is the time, should 
return a value for $b$ that is consistent with zero 
(i.e. the error is much greater than the value of the coefficient itself)
For all models, only S2 and S3 have barely significant coefficients $b$,
albeit marginally ($p$-value = 0.08 and 0.06, for S2 and S3, respectively), 
reinforcing the uniformity test indicated above. 
The CV formation rate in model S2 is not exactly constant. 
The other models are compatible with the hypothesis of a constant rate.

The result of this last test is shown in Fig. \ref{Fig03.2}, which
shows the number of CVs formed in intervals of 1 Gyr. In this figure,
the blue line is the linear regression, the gray band is the 
95\% confidence interval around the linear regression (which indicates 
the error in the fitting procedure), the error
bars correspond to Poisson errors ($\sqrt{n}$), and the
dashed line is the average rate (assuming a constant rate, i.e. $b=0$). 
Note that straight lines within the gray band are compatible with the data, 
to within the fitting errors. Additionally, all dashed lines (average rates) are inside 
the gray band, which indicates that a constant CV formation rate is coherent
for all models, although models S2 and S3 barely fit this criterion, 
since their b values are barely non-zero ($p$-values of 0.08 and 0.06,
respectively, but still higher than the classical 0.05 for 
statistical significance).

We emphasize that the one-sample KS test and the linear regression fitting 
are different procedures, which leads to different results. 
The former should be somewhat more precise, since the latter depends on the 
choice of binning for the histogram counts. However, given the differences in the results, 
it seems reasonable to assume that the CV formation rate is nearly constant.

To summarize, the WD-MS binary formation rate
shows an initial burst, followed by a roughly smoothly decreasing
rate.  The CV formation rate, on the other hand, has an initial burst, 
followed by a nearly constant formation rate.  This difference is due to the 
time-delay between WD-MS binary and CV formation. This suggests that we should 
expect the same CV formation rate in all types of clusters.
However, only by analyzing more models can we better define 
any dependence of the WD-MS binary and CV formation rates on the host 
cluster properties and evolution.

\subsubsection{Comparisons with \citet{Ivanova_2006}}
\label{GC-EVOL-Ivanova}

Now that the dependences of the WD-MS binary and CV formation rates on the 
cluster properties have been 
described, we compare our general picture with the findings
of \citet{Ivanova_2006}.

In their Fig. 8, they show the rate of occurrence of the
last major dynamical event or CEP. Notice that this rate corresponds 
closely to our WD-MS binary formation rate. They show that
the occurrence of CEPs is mostly within the first 
$\sim$ 3 Gyr of cluster evolution, and the occurrence of dynamical events 
is distributed almost uniformly in time. 

As discussed in Section \ref{GC-EVOL-cv}, we find 
in our simulations an enhancement in the formation rate 
of WD-MS binaries in the first few Myr of 
the cluster evolution, for both PCEBs and dynamically
formed WD-MS binaries.  This is followed by a smoothly decreasing
rate.  Both episodes can be explained by the evolution in the turn-off mass, 
and the interplay between formation and destruction of WD-MS
binaries.

The above differences in our findings can be explained by the differences inherent 
to both approaches. 
In our case, we model CV formation throughout the entire cluster, 
whereas they consider only formation in the core. Additionally, their clusters
are frozen in time, while ours evolve dynamically.  As a result, 
they might be missing some of the CVs formed in our simulations, either at 
early or late times, which are of course responsible for the above-mentioned features.

In their Fig. 9, they show the initial appearance of CVs (i.e., the onset of mass 
transfer) in their simulations. This corresponds to our CV formation rate.
They find for CVs formed from a CEP that the CV formation
rate is roughly constant in time.  For dynamically
formed CVs, they find an increase in the formation rate after $\sim$ 7 Gyr.
In our simulations, we find that CVs formed through all channels
show an enhancement in the rates at the beginning of the cluster evolution, followed 
by a nearly constant rate after $\sim$ 1 Gyr.  This is accounted for by the time-delay 
between WD-MS binary formation and CV formation.  
Additionally, we applied a two-sample KS test, and concluded that
we do not have enough evidence to reject with 99 per cent
confidence (or more) the null hypothesis that the CV formation 
rate is unique for all GCs.

In general, the overall results of the two studies corroborate, especially 
given their respective limitations. Yet to come detailed investigations
could define with more confidence the relation between the cluster 
properties and the CV formation rate.

\subsection{The formation-age population}
\label{FAP}

\begin{figure*}
   \begin{center}
    \includegraphics[width=0.362\linewidth]{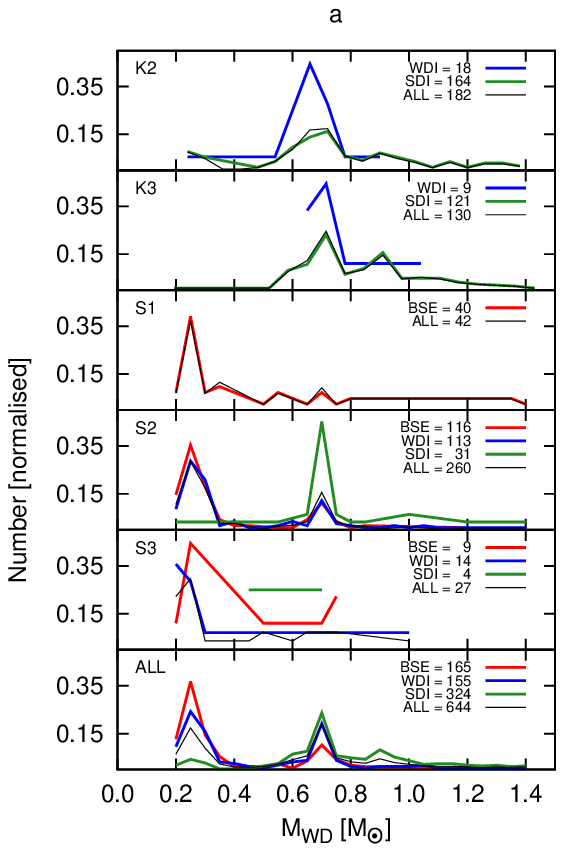} 
    \includegraphics[width=0.315\linewidth]{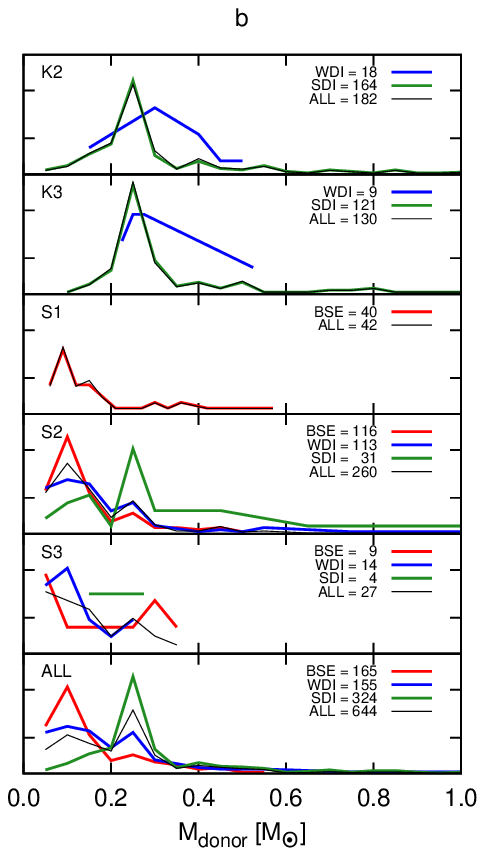} 
    \includegraphics[width=0.315\linewidth]{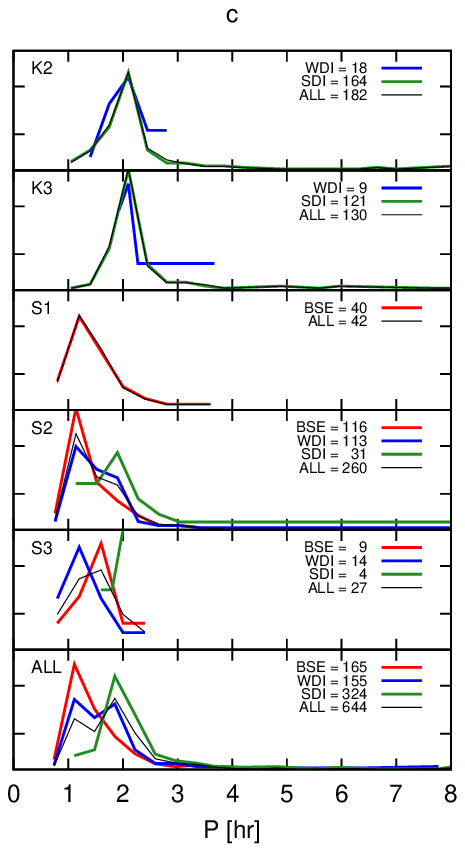} 
    \end{center}
  \caption{WD mass (left-hand panel), donor mass (middle panel) and period 
(right-hand panel) distributions associated with
CVs that survive up to the present-day, shown at their formation times in the six models.
Keys and the bottom row are similar, as in Fig. \ref{Fig01}. 
We can see a clear distinction between CVs formed under the influence
of strong dynamical interactions (SDI group) and CVs formed without any influence
from dynamical interactions (BSE group). CVs in the BSE group have predominantly
low-mass WDs and donors, and extremely-short periods. CVs in the SDI group have
higher WD and donor masses, and longer periods. CVs formed via the weak 
influence of dynamics (WDI group) have properties similar to CVs in the BSE and 
SDI groups. For more details, see Section \ref{FAP}.}
  \label{Fig04}
\end{figure*}

The WD and donor mass, and period distributions of the CVs 
at the moment they are formed are shown in Fig. \ref{Fig04}.
In what follows, we focus more on the bottom row of each
panel, which corresponds to our `aggregated/average' cluster.

\subsubsection{Primary Mass}
\label{fap_primary_mass}

We see from the WD mass distribution (\ref{Fig04}a) that there is a
clear separation between dynamically formed CVs and CVs formed
without any influence from dynamics. For the CVs in the BSE group,
the peak is $\lesssim$ 0.25 ${\rm M_\odot}$, which means that 
those CVs formed from CEPs without any influence from dynamics
typically have He WDs rather than C/O WDs. Additionally, there is 
a smaller peak at $\sim$ 0.7 M$_\odot$, which is due to 
massive WDs formed early on in the cluster evolution.

On the other hand, CVs in the SDI group have 
more massive WDs, with a peak at around 0.7 ${\rm M_\odot}$.
This is explained via the role of exchanges in the CV 
formation process (Section \ref{fap_exchange}).  
There are other small peaks in the 
distribution as well, at $\sim$ 0.25 
(caused by the evolution of low-mass MS stars) and $\sim$ 0.9 M$_\odot$
(caused by the evolution of high-mass MS stars).

CVs in the WDI group have roughly 
the same relative numbers of massive and light WDs. This is due
to the net effect of weak dynamical interactions. If the net
effect is great, then more massive systems can be 
bumped into the required range in parameter space to form CVs.
On the other hand, if the net effect is weak, then these 
interactions are unable to change drastically the binary
properties, such that the CV will have similar properties to those 
found in the BSE group. 

CVs in the WDI group in the 
Kroupa models are similar to those in the SDI group.  This 
indicates that in these clusters, the net effect of 
weak dynamical interactions is strong.  For the
Standard models, most CVs in the WDI group
behave analogously to those in the BSE group (the net effect of weak dynamical
interactions is weak).  However, we see a few for which the 
cumulative effect is strong.  These objects have similar features as the
CVs in the SDI group.

Importantly, the WD mass distribution
at the time of formation is very similar to that at the present-day.  
This is because BSE assumes that only 0.001 per cent of the
accreted matter is, in fact, accreted by the WD 
\citep[][Eq. 66]{Hurley_2002}. Considering
the low mass transfer rates of these systems (short-period CVs),
we should not expect too much of a difference in the WD masses.
This is a reasonable expectation since, due to nova eruptions,
most of the accreted material is indeed expelled (but not all).

\subsubsection{Secondary Mass}
\label{fap_secondary_mass}

As for the donor mass, we see the same behaviour
for the WD mass, i.e. a clear separation between 
dynamically formed CVs and CVs formed without any influence 
from dynamics. Dynamically formed CVs have more massive
donors {\it at the formation time}, relative to CVs not influenced strongly
by dynamics, which is in good agreement with the findings
in \citet{Shara_2006}. 

This is once again associated with exchanges. From the
second panel of Fig. \ref{Fig01}, we see that most CV progenitors 
in the SDI group have mass ratios greater than $\sim$ 0.2.  This 
is the limit for CV formation through pure stellar 
evolution (Section \ref{pp_mass_ratio}).  These binaries 
have to have at least one binary component replaced in order to become CVs. 
After the exchange, the resulting binaries are composed of a massive star 
(either a WD or an MS star) and an MS star with mass $\lesssim$ 0.5
${\rm M_\odot}$.  These MS stars are future CV donors.

On the other hand, CVs in the BSE group have donor masses
$\lesssim$ 0.2 ${\rm M_\odot}$, because their progenitors had
such a small range of masses (see mass ratio and primary mass distributions
in Fig. \ref{Fig01}).

It is important to note that we do not find any difference 
in the donor mass in the BSE group relative to what is found for the dynamics groups 
at 12 Gyr \citep[][see their Fig. 2]{Belloni_2016a}.  This is the result of 
CV evolution (mass loss).

In the end, most CV donors have masses $\lesssim$ 0.35 ${\rm M_\odot}$,
which implies that they are born either as short-period CVs or 
as gap CVs.

\subsubsection{Mass Ratio and Period}
\label{fap_mass_ratio}

Beginning with the mass ratio distribution, almost all CVs have mass
ratios less than 1 (only one CV in model S2 has q > 1).  This means 
that they are thermally stable. However, some of them
could be dynamically unstable (see Section \ref{fap_discussion}).
Additionally, most CVs have mass ratios between 0.2 and 0.6, 
and there is no clear distinction among the BSE, WDI and SDI groups.

As already inferred from the
donor masses, most CVs have periods $\lesssim$ 3 h, which
means that they are short-period CVs or gap CVs. We found
in our simulations that the gap is apparently missing.  

In general, dynamically formed CVs have longer
periods, since they have heavier donors. CVs in the 
BSE group have very low mass donors, which leads them
to have very short periods. This comes from the fact that
CV donors fill their Roche lobes, which produces 
a rough relation between donor mass and CV
period \citep[e.g.][see section 2.1]{Knigge_2011ASPC}.

\subsubsection{Discussion on the Formation-Age Population}
\label{fap_discussion}

We begin this section with a few comments regarding potential 
biases that could affect any interpretations of our results.  
First, with respect to WD mass, it might be that most of 
the low-mass CVs (composed of He WDs) are dynamically
unstable. According to \citet{Schreiber_2016}, if the
strength of consequential angular momentum loss 
is inversely proportional to the WD mass, then most CVs with WD masses
less than $\sim$ 0.5 ${\rm M_\odot}$ are dynamically unstable.
Note that we use the term "consequential angular momentum loss" to refer 
to AML that is a consequence of mass-loss.
The motivation for such a functional form comes from the fact
that a He WD has never been observed in a CV in the Galactic field, although
such WDs are observed in Galactic detached systems \citep{Zorotovic_2011}.
The primary mechanism thought to be responsible for such a dependence for the WD mass
is nova eruptions \citep{Nelemans_2016,Schreiber_2016}. 
The frictional angular momentum loss produced by novae depends strongly 
on the expansion velocity of the ejecta \citep{Schenker_1998}. For low-mass WDs, the
expansion velocity is small \citep{Yaron_2005}, and this leads 
to a strong angular momentum loss by friction.  This, in turn,
makes CVs dynamically unstable.

We emphasize that the apparent absence of the period 
gap in our simulations might be an intrinsic issue
in the BSE code. This is because, even without dynamics,
we cannot reproduce the period gap during the evolution
of long-period CVs with BSE. This is most likely connected 
with the remaining improper implementations in BSE  
(e.g. tides and spins).

\subsection{From the formation-age to the present-day}
\label{FAP-PDP}

In this section, we briefly discuss CV evolution from
the time of their birth until a cluster age of 12 Gyr. Special attention is given
to one unique case that suffered a strong dynamical interaction
after CV formation. As a general characteristic, 
CVs are hard enough to avoid any kind of strong encounter.  However, 
some can still undergo weak interactions, which causes a minuscule change in 
eccentricity (see \citet{Leigh_2016} for more details regarding the interruption of 
binary mass transfer due to dynamical interactions).

\subsubsection{CV evolution without dynamics}
\label{cv_evolution_without}

For the most part, CV evolution without dynamics is as we would expect.  
That is, a CV evolves toward shorter periods up to the moment its 
donor changes its structure, and becomes an H-rich degenerate object. 
At this point, the CV begins to evolve toward longer periods.

As already mentioned, BSE cannot reproduce
the period gap. This is apparent upon analyzing CV 
evolution for the long-period CVs.  In BSE, there
is a decrease in the rate of angular momentum loss when the donor
reaches a mass of $\sim$ 0.35 ${\rm M_\odot}$ 
\citep[][section 2.4]{Hurley_2002}, because magnetic 
braking ceases to act.  At this point, in reality, the donor 
would have time to restore thermal equilibrium and stop overfilling its
Roche lobe.
However, in BSE, tides (in combination with spins) keep the binary 
locked into mass transfer, which causes the period to increase in response.

From a statistical point of view, 
the aforementioned issues should not drastically affect our results.  This is 
because our main objective in this paper is not to model particular CVs, 
but rather to establish an overall dynamical picture for CV formation and evolution 
in globular clusters.

\subsubsection{CV evolution with dynamics}
\label{cv_evolution_with}

Next, we discuss CV evolution altered 
by some kind of dynamical interaction.  \citet{Shara_2006}
described one such case found in their simulations.  This CV had its 
evolution accelerated due to dynamical interactions.

In our simulations, we found only one similar case (out of 644 CVs), 
in model K2. Unfortunately, model K2 has only dynamically
formed CVs, so that we cannot make a comparison with
a field-like CV population, as done by \citet{Shara_2006}.

This CV in model K2 is not the result of an exchange. Instead, it is the result of
a sequence of dynamical interactions that decreased its progenitor's orbital
period, causing the outcome of the CEP to be a short-period
binary. The CV formed at $\sim$ 200 Myr.  After its formation,
it was unaffected by dynamical interactions,
up to $\sim$ 10 Gyr, at which point this CV interacted with a WD-WD binary.
At this moment, the CV 
is already far past period bounce, and has a donor mass of
$\sim$ 0.04 ${\rm M_\odot}$ and a period of $\sim$ 3 h. The WD-WD binary 
initially had a period of $\sim$ 600 hr. After the interaction, 
as expected, the CV became harder, and the WD-WD binary softer. 
This hardening led to a CV with an orbital period of $\sim$ 1 h, 
causing an acceleration in its subsequent evolution due to an enhancement 
in the rate of mass loss, and a decrease in donor mass (to a value 
$\sim$ 0.025 ${\rm M_\odot}$). Within a fraction of a Myr 
after the interaction, the CV established a stable mass transfer rate 
with a period of $\sim$ 4 h.

Even though both CVs (in our simulations and in their simulations)
corroborate with respect to the pattern, we expect
this behaviour to be rare in GCs \citep{Leigh_2016}.  This is because CVs are very dynamically 
hard, and of low mass, such that the impact parameter for interactions is small.  
It follows that the probability of an interaction occurring is very small.  
Only one case in our investigation and one case in the \citet{Shara_2006} study 
were found, suggesting that overall dynamics should only very rarely directly 
affect CV evolution.

On the other hand, as discussed in Section \ref{fap_acceleration},
pre-CVs are more likely to interact.  Consequently, this is a more likely channel to 
eventually affect CV evolution, whether it be in the form of an 
acceleration or a retardation.  But we emphasize that the influence of dynamics typically 
affects CV progenitors, not CVs directly.  Dynamics acts indirectly, by affecting the 
progenitors.

\subsection{Dependence on initial binary fraction}
\label{BIN_FRAC}

In this section, we test to what degree the results presented in 
this paper (relative to \citet{Belloni_2016a}) depend on our 
assumptions for the initial binary fraction, for both adopted 
IBPs (high for Kroupa IBP and low for Standard IBP).  To this end, 
we ran 3 additional models with the same initial cluster properties 
as in the S1, S2 and S3 models, but with an initial binary fraction 
of 95 per cent.

For these runs, models S1, S2 and S3 have semi-major 
axis distributions extending initially up to 50 AU.  We emphasize 
that by increasing the initial binary fraction, we must also increase 
this maximum semi-major axis. Otherwise, the energy generated via 
single-binary encounters catalyzed by the presence of such a large 
population of very hard binaries would be incredibly high, and 
present-day clusters would have much higher binary fractions than observed.  
Consequently, in the Standard models with initial binary fractions of 
95 per cent, the initial semi-major axis distribution extends to $10^4$ AU.  
After 10--12 Gyr of evolution, however, the cluster binary fraction 
(including all binary masses) has been reduced to around 30, 20 and 10 
per cent, in order of increasing cluster density. Most of these binaries 
are of low-mass, remain in the outskirts of the system, and do not affect 
the rate of energy generation due to binary interactions or the overall 
cluster evolution.

The resulting CV properties characteristic of these models are similar to
 models S1, S2 and S3.  The main difference between these sets of models 
is the total number of CVs that are formed.  Models with high initial binary 
fractions form more CVs than models with low initial binary fractions.
But their CVs still form at an approximately constant rate due mainly to 
binary stellar evolution, are composed of mostly low-mass WDs, tend to 
form with short periods (< 2h), and are often period bouncers at the present-day.

\subsection{CV properties and dependences on their age}
\label{AGE}

Having described the main features characteristic of the present-day
CVs in our six models, from the time of cluster birth up to an age of 12 Gyr, 
we now turn our attention to a more general issue concerning the time 
evolution of the properties of CV populations, before addressing 
CVs that do not survive until the present-day (Section \ref{DEST}).

\begin{figure*}
   \begin{center}
     \includegraphics[width=0.98\linewidth]{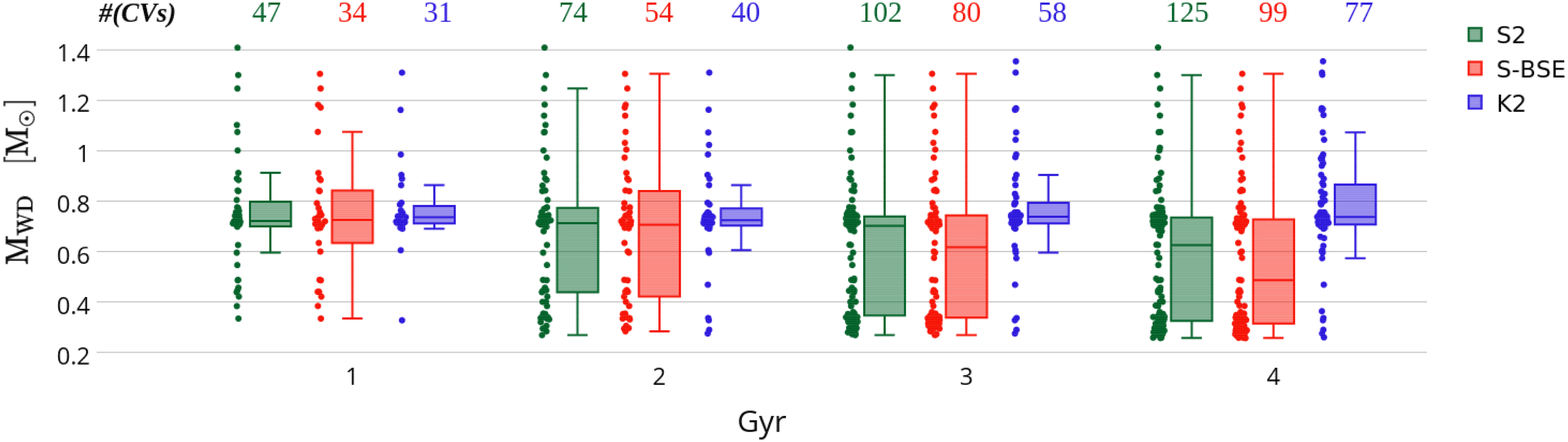} 
     \includegraphics[width=0.98\linewidth]{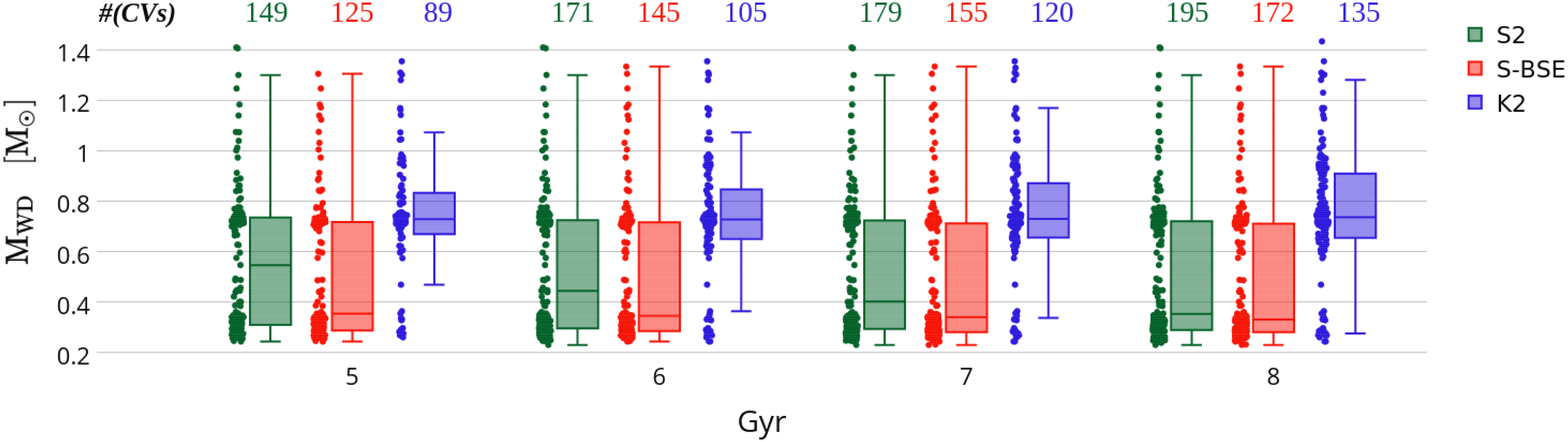} 
     \includegraphics[width=0.98\linewidth]{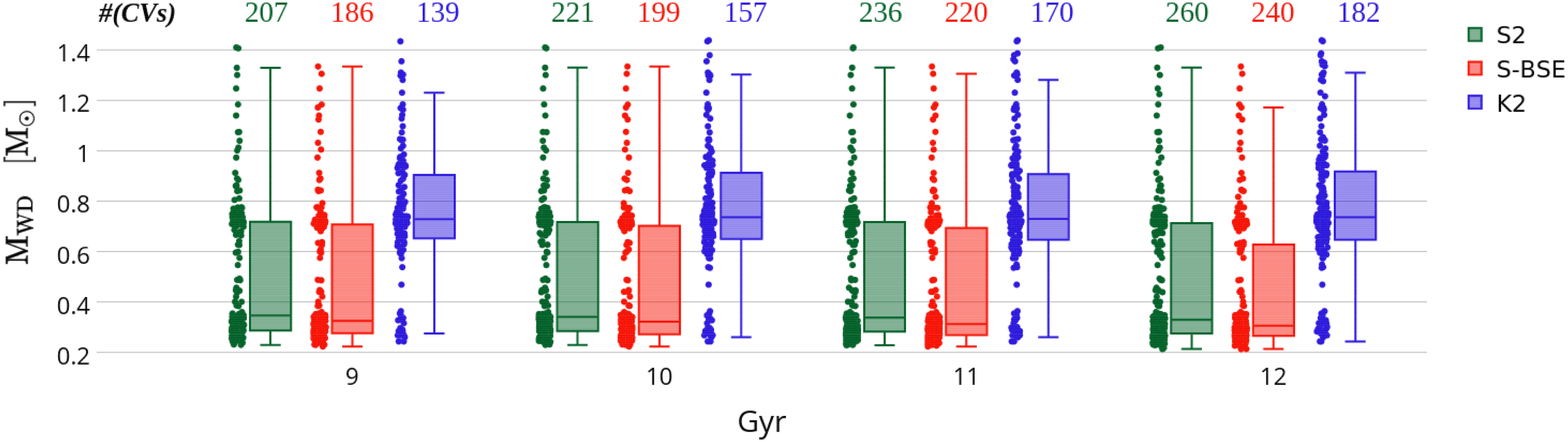} 
   \end{center}
  \caption{Evolution of the WD mass distribution in time for 
all CVs in 2 out of the 6 models. Models K2 and 
S2 show the results for CVs evolved by MOCCA (i.e. they are associated 
with a cluster environment). Model S-BSE is related to 
model S2, but evolved using BSE alone (i.e. it is 
related to a field-like environment, without dynamics).  
Each box corresponds to the region between the first and third quartiles,
with a horizontal line at the median value. Hence, 25 per cent of the data
belongs to the region below the lower edge of the box, 25 per cent of the data
belongs to the region above the upper edge of the box, and the box
corresponds to 50 per cent of the data (interquartile region).  
Above and below the boxes there are two vertical lines, called the whiskers.
The line above the box ends at a value 
which is 1.5 times the interquartile range (i.e. the vertical
height of the box) from the upper edge of the box.
Similarly, the line below the box ends at a value which
is 1.5 times the interquartile range, from the lower edge.
Each whisker is truncated at a WD mass value belonging to
a particular data point. Since there may be no data point whose
value is exactly 1.5 times the interquartile distance, 
the whisker may be shorter than its nominal range. This is
the case for the line below the boxes.  
For some boxes, the lower edge is very close to the end
of the vertical line below the box. This is because
there are no data points below this. Any points that lie 
outside the range of the whiskers are considered outliers.
Each cloud of points to the left of each box represents 
the distribution of WD masses, and above each box the number 
of CVs in the models at a particular time is given. 
Notice that model S2 contains more high-mass WDs than model
S-BSE due to dynamical exchanges that put massive WDs into the CV population.
Regardless, both models are dominated by low-mass WDs after $\sim$ 4 Gyr.  
Model K2 is dominated by massive WDs throughout the cluster evolution, although
a non-conspicuous peak is observed for low-mass WDs. For more details, 
see Section \ref{time_primary_mass}.}
  \label{Fig06}
\end{figure*}

\begin{figure*}
   \begin{center}
     \includegraphics[width=0.98\linewidth]{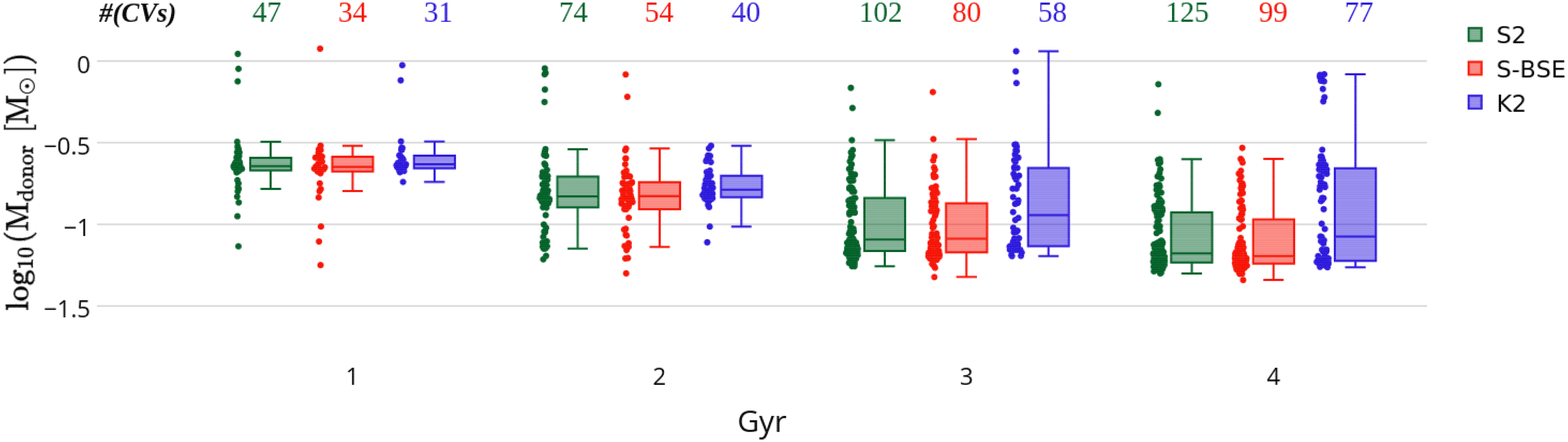} 
     \includegraphics[width=0.98\linewidth]{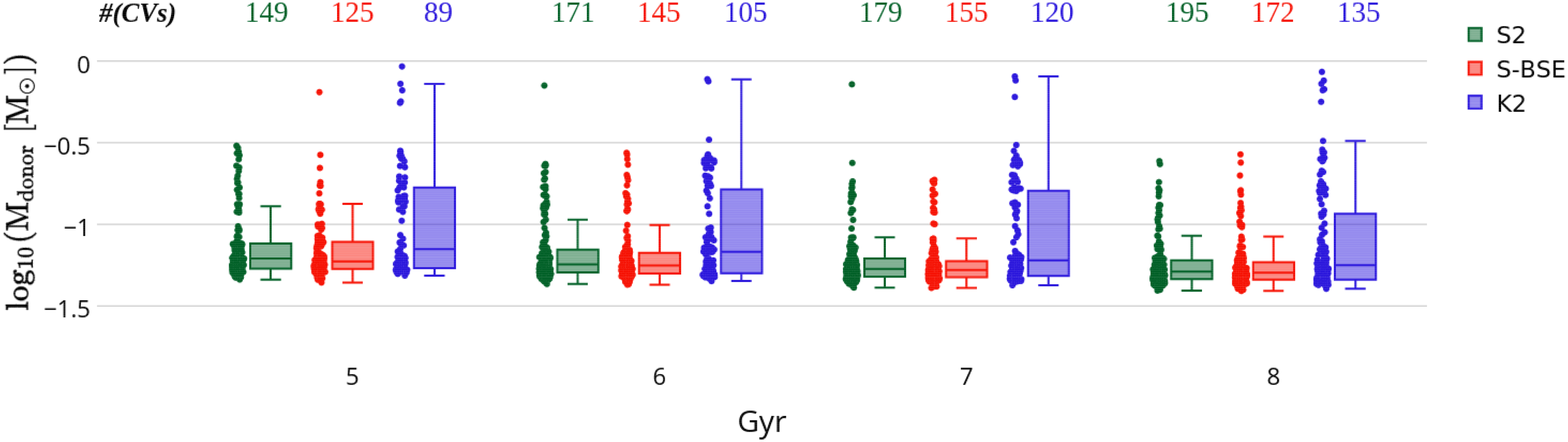} 
     \includegraphics[width=0.98\linewidth]{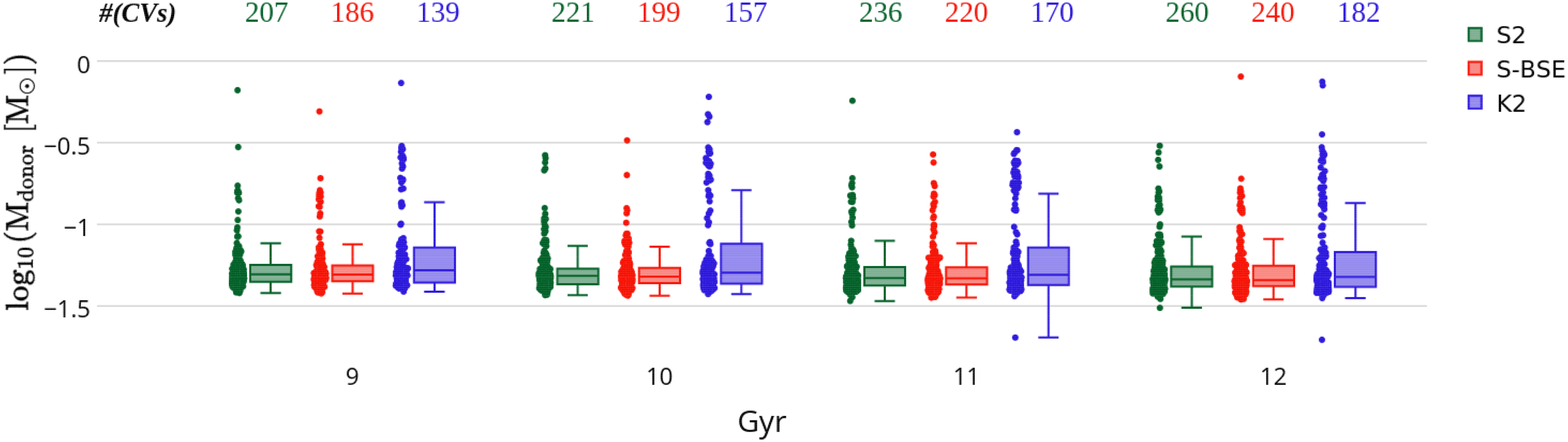} 
   \end{center}
  \caption{Evolution of the donor mass distribution in time for 
all CVs in 2 out of the 6 models. The notations are the same 
as in Fig. \ref{Fig06}. Notice that the donor masses decrease
with time not only via the addition of new systems, but also
due to CV evolution (mass transfer).  
Model K2 has slightly more massive donors than models S2 and S-BSE.
For more details, see Section \ref{time_secondary_mass}.}
  \label{Fig07}
\end{figure*}

One important property of CVs, in the Galactic field 
or in GCs, is that CV populations as a 
whole change properties as 
they age. This is important for comparing field and cluster CVs, since cluster 
CVs might be up to 4 times older than field CVs.

Note that CVs are created continuously in time 
(see Fig. \ref{Fig03}d).  Hence, the properties of 
CV populations should change in time, due both to CV evolution
and the constant addition of new CVs to the population. 

In what follows, we describe the main changes in the 
CV population properties as a function of time. We concentrate
on changes in the WD (Fig. \ref{Fig06}) and donor mass (Fig. \ref{Fig07}) distributions.  
In both figures, each box corresponds to the region between the first and third quartiles,
with a horizontal line at the median value. Hence, 25 per cent of the data
belongs to the region below the lower edge of the box, 25 per cent of the data
belongs to the region above the upper edge of the box, and the box
corresponds to 50 per cent of the data (interquartile region).
Above and below the boxes there are two vertical lines, called the whiskers.
The line above the box ends at a value 
which is 1.5 times the interquartile range (i.e. the vertical
height of the box) from the upper edge of the box.
Similarly, the line below the box ends at a value which
is 1.5 times the interquartile range, from the lower edge.
Each whisker is truncated at a value in mass belonging to
a particular data point. Since it can occur that no data point has a 
value of exactly 1.5 times the interquartile distance,
the whisker may be shorter than its nominal range. Additionally,
a cloud of random points is shown to the left of each box, which corresponds
to the underlying distribution.

In Fig. \ref{Fig06} and Fig. \ref{Fig07}, two clusters were considered, namely
S2 and K2. In order to perform comparisons to a field-like population, 
we added an additional population similar to model S2, but evolved using BSE alone (i.e. 
without dynamics). This third model is named in the figures 
as S-BSE. Importantly, the other K and S 
clusters show similar features. This is the reason we show
only these two clusters in the figures.

\subsubsection{WD Mass}
\label{time_primary_mass}

Fig. \ref{Fig06} shows the time evolution of the WD mass distribution
for CVs.  Note that for model K2 (having only dynamically formed CVs), 
the WD median mass stays roughly constant in time ($\sim$ 0.7 ${\rm M_\odot}$). 
This illustrates what has already been said about the
role of exchanges (Section \ref{fap_exchange}), which put more high-mass
WDs into the CV population.  We also see a second inconspicuous
peak for low-mass WDs around 0.3 ${\rm M_\odot}$, which originates from 
CVs affected by weak dynamical interactions, and becomes more pronounced
at $\sim$ 8 Gyr. Nevertheless, it is clear from the figure that model 
K2 is dominated by high-mass WDs. This is not true for model S2.

In models S2 (cluster CVs) and S-BSE (field-like CVs), 
we see a clear drop in the WD median mass, starting 
at $\sim$ 3 Gyr. This is caused by the continuous addition of 
low-mass WDs in to the CV population, due to the 
evolution of MS stars whose masses are 
$\lesssim$ 2 M$_\odot$. In the beginning
(up to $\sim$ 1 Gyr), both models are dominated by high-mass
WDs. After that, we see the addition of more low-mass WDs than
high-mass WDs. We also see that the cluster CVs (model S2) have 
slightly more high-mass WDs than their field-like counterparts (S-BSE)
throughout the evolution, due to a few CVs that are formed due 
to strong dynamical interactions.
Furthermore, from the distributions near the boxes, we see that
for these two models, the WD mass distribution becomes bimodal
at $\sim$ 3 Gyr, although the peak for high-mass WDs ($\sim$
0.7 M$_\odot$) is much less pronounced than the peak for
low-mass WDs ($\sim$ 0.3 M$_\odot$).

Finally, we point out two main features in Fig. \ref{Fig06}; 
(i) one associated with the time evolution of the WD mass distribution,
and (ii) the other with the differences among the models.

(i) We expect a clear drop in
the median WD mass over time due to the addition of low-mass WDs. However, 
if the cluster is dominated by dynamically formed CVs,
we expect roughly no time evolution of the median value. 

(ii) We observe an extension 
of the interquartile range in the early evolution for all models.  
However, for models S2 and S-BSE, a decrease is seen instead during 
the late-time evolution, which is caused by the predominance of low-mass
WDs.

\subsubsection{Donor Mass}
\label{time_secondary_mass}

From Fig. \ref{Fig07}, we see a drop in the median value of the 
donor mass distribution for all models as time goes on.  However, 
the drop is less severe for dynamically formed CVs 
(model K2). This indicates that CV populations in all models 
tend toward having extremely-low-mass donors at late 
times, regardless of the influence of dynamics or the formation 
channel.  
More specifically, dynamically formed CVs tend to have
more high-mass donors, relative to CVs formed without any
influence from dynamics (compare models K2 and S-BSE). Furthermore,
model S2 (a dynamical environment) has slightly more 
high-mass donors than model S-BSE (a field-like 
environment). This is because the greater is the WD mass, 
the greater can the donor mass be in order for the 
stability limit to remain respected. Since dynamically formed CVs tend to have
greater WD masses, we expect greater donor masses as well.  
In the beginning, the boxes in Fig. \ref{Fig07} are 
very narrow and have mainly high-mass donors, for all models.
After $\sim$ 3 Gyr of evolution, the median drops 
below 0.1 M$_\odot$ in models S2 and S-BSE.  In 
model K2, however, the median remains above this value. 

At $\sim$ 4 Gyr,
the median mass for all models is below 0.1 M$_\odot$. After that, 
the distributions for models S2 and S-BSE 
become completely dominated by extremely-low-mass donors, 
with a huge peak around the first quartile and a long tail
toward higher masses. This feature remains up to a cluster age of 12 Gyr. 
The long tail is caused by the constant 
addition of CVs to the population, and the huge peak
by CV evolution.

On the other hand, model K2 maintains a bimodal distribution
from $\sim$ 4 Gyr up to $\sim$ 7 Gyr. At this point ($\sim$ 7 Gyr), 
the CV population in this model has more than 50 per cent of the entire
CV present-day population (at 12 Gyr). After 7 Gyr,
the CV population has a similar feature to the other models:
a huge peak around the first quartile, and a long tail
towards higher masses. Notice that, at $\sim$ 8 Gyr,
the CV population in model K2 is dominated by extremely-low-mass donors,
and the addition of more massive donors is not able
to further maintain the bimodal shape of the distribution.

Interestingly, we can clearly see in the boxes (at 11 and 12 Gyr) for model K2 the
CV mentioned in Section \ref{cv_evolution_with}, which had its evolution accelerated
by a strong dynamical interaction. Note that, in the figure, such a CV donor
appears as an isolated point below $\log_{10}(M_{\rm donor}) = -1.5$, which is a
visualization of the effects of strong dynamical interactions that accelerate
CV evolution. Although such objects are quite interesting from a dynamical
perspective, they are rare as already pointed out and probably part
of the faintest population of CVs in GCs, which makes them unlikely
to be observed.

\subsubsection{Discussion on the age-dependence of CV properties}
\label{time_period}

Importantly, it is not just CV evolution that causes 
the decrease in the median donor mass in the overall population in models
S2 and S-BSE. Additionally, 
more low-mass CVs are added to the population as time goes on. Again, 
this is associated with the turn-off 
mass evolution, which makes low-mass WDs appear more frequently at late times.
We showed in Section \ref{GC-EVOL-cv} that the CV formation rate
is driven by the WD formation rate in the cluster, and the time delay
needed for WD-MS binaries to become CVs. We further showed that
the time delay is small in comparison with the cluster lifetime.
This implies that we should, indeed, expect the addition of
more low-mass CVs to the entire CV population, relative to higher mass
CVs.

On the other hand, in model K2 (i.e. for dynamically formed CVs),
we find that the WD mass distribution is dominated by high-mass
WDs. Additionally, even though the newly added CV donors in model K2
are more massive than the newly added CV donors in models S2 and S-BSE,
we still find that after $\sim$ 8 Gyr, the CV population is predominantly 
composed of extremely-low-mass donor, which is caused mainly 
by mass loss due to CV evolution.

Having described the overall picture for the age-dependence
of CV properties, we now turn to a discussion of the 
implications of these features for CV populations.

The donor mass evolution is very important for determining 
the CV status at any given time. In the case of cluster CVs,
we clearly see from Fig \ref{Fig07} that the majority of 
CVs have extremely-low-mass donors after 12 Gyr 
of cluster evolution. This means that they are
much fainter than the Galactic field CVs as a whole, if one assumes
that the star formation rate in the Galaxy is roughly constant and
bears in mind that star formation in GCs took place only 
when the clusters formed. Thus, we expect that cluster
CVs should have donors with lower masses than the observed Galactic field
CVs.

This brings serious restrictions to any comparison between 
{\it predicted} cluster CVs and {\it observed} field CVs. 
The old field CVs are still missing, although a few of them have now 
been detected  \citep[e.g.][]{Littlefair_2006,Santisteban_2016}.
Regardless, the majority of the observed CVs in the field seem to be
much younger than GC CVs. 

\citet{Ak_2015} inferred kinematic ages in a sample of field CVs and 
concluded that 94 per cent of CVs in the solar neighbourhood belong to 
the thin-disc component of the Galaxy. Mean kinematical ages of 
3.40 $\pm$ 1.03 and 3.90 $\pm$ 1.28 Gyr were found for the non-magnetic thin-disc 
CVs below and above the period gap, respectively. Some GC CVs can be up to 4 
times older than the observed field CVs. This means that comparing GC CVs 
and field CVs should be done carefully, especially if one considers {\it predicted}
GC CVs and {\it observed} field CVs.

\subsection{Destroyed CVs}
\label{DEST}

Thus far, we have described CVs formed in 
clusters that survive until the present-day. In this
section we deal instead with CVs that do not survive, and are 
not found in the present-day clusters. These are called
`destroyed' CVs\footnote{The reader should keep in mind that
the term `destroyed' used here is not necessarily associated
with a destructive process {\it stricto sensu}. The term
adopted here has a more general sense: CVs that are created
during the cluster evolution but do not survive in the cluster up to the
present-day, which can occur for many reasons.}, in the sense that they are formed but do 
not survive until a cluster age of 12 Gyr.

\citet[][see section 3.5.1]{Belloni_2016a} describe the three
main destruction channels associated with cluster CVs.  These 
are: (i) destruction due to binary stellar evolution
(the CV stops being a CV due some evolutionary process
and without any influence from dynamical interactions); 
(ii) destruction
due to escape (the CV life in the cluster is interrupted because the
CV escaped the cluster, independent of whether or not the CV remained 
a CV post-escape); and 
(iii) destruction due to dynamical interactions (the CV death was caused by 
a dynamical interaction, but the outcome of this interaction remains 
in the cluster).

In what follows, we describe the 
main features associated with the above destruction channels.
The number of CVs that do not survive until 12 Gyr, 
separated according to their main formation and destruction channels, 
are given in Table \ref{Tab2}. 

\begin{table}
\centering
\caption{Number of CVs that are formed during the cluster evolution,
but are not present in the PDP, 
separated according to their main formation and destruction channels.
We also indicate the number of CVs for which exchanges are the main
dynamical process involved in their formation.}
\label{Tab2}
\begin{adjustbox}{max width=400px}
\noindent
\begin{threeparttable}
\noindent
\begin{tabular}{l|c|c|c|c|c|c|c|c}
\hline
Model & \multicolumn{4}{c}{Formation Channels} & \multicolumn{3}{c}{Destruction Channels} & Total \\
\hline
 & BSE \tnote{a} & WDI \tnote{b} & SDI \tnote{c}  & Exchange & DBSE \tnote{d} & DESC \tnote{e} & DDI \tnote{f} &  \\ 
\hline\hline
K1 &   45 &  43 &   4 &    1 &   89 &   3 &   0 &   92  \\ \hline
K2 &   29 & 111 & 272 &  144 &  398 &   3 &  11 &  412  \\ \hline
K3 &   18 &  69 & 211 &  102 &  277 &  10 &  11 &  298  \\ \hline
S1 &  114 &   2 &   0 &    0 &   97 &  19 &   0 &  116  \\ \hline
S2 &  169 & 131 &  24 &   11 &  324 &   0 &   0 &  324  \\ \hline
S3 &    7 &  23 &  18 &   14 &   33 &  14 &   1 &   48  \\ \hline 
\hline 
\end{tabular}
\begin{tablenotes}
       \item[a]  Binary Stellar Evolution
       \item[b]  Weak Dynamical Interaction
       \item[c]  Strong Dynamical Interaction
       \item[d]  Destruction due to Binary Stellar Evolution
       \item[e]  Destruction due to Escape
       \item[f]  Destruction due to Dynamical Interaction
\end{tablenotes}
\end{threeparttable}
\end{adjustbox}
\end{table}

\subsubsection{Unstable CVs}
\label{unstable_CVs}

From Table \ref{Tab2}, we see that 
the most prominent destruction channel is unstable mass
transfer that leads the CV to merge, which is in good agreement
with \citet{Ivanova_2006} and \citet{Shara_2006}. This happens because
the WD cannot stably burn the accreted material \citep{Nomoto_2007}, which
causes even higher mass transfer rates, and, subsequently, 
a CEP followed by a merger.

Considering all six models, more than 95 per cent of 
the destroyed CVs are destroyed because of unstable mass
transfer. In the Kroupa models, they are mainly formed
due to strong dynamical interactions.  In the Standard
models, they are mainly formed due purely to a CEP.

The reason for the unstable mass transfer is the high mass
of the donor, which is much greater than the WD mass. This
implies that the CV mass ratio is much larger than unity.
The critical mass ratio that separates stable and unstable
mass transfer can be obtained by equating the adiabatic 
mass-radius exponent and the mass-radius exponent of the donor 
Roche-lobe. Such a critical mass ratio is of order unity 
for almost all donors, if one assumes the classical version
of the angular momentum loss prescription 
\citep[][for more details]{Schreiber_2016}. 
Thus, CVs with very massive donors 
($M_{\rm donor}$ $\gtrsim$ $M_{\rm WD}$) will
probably undergo another CEP, and merge.

It is important to note that few CVs with $M_{\rm donor}$ $\sim$ $M_{WD}$
will evolve towards stable mass transfer. In all six models, we detected only 
one such case. CVs undergoing unstable mass transfer live 
only for a few Myr.

\subsubsection{Escaping CVs}
\label{escaping_CVs}

Anther possible mechanism for destroying CVs is their 
escape from the cluster, due either to relaxation or direct (strong) dynamical interactions.  
Escape due to relaxation takes place gradually, and the CV probably
remains as a CV after the escape. On the other hand, escape
due to dynamical interactions is a more violent process; 
the binary typically has its orbital parameters changed considerably, and
the CV is probably no longer a CV after the escape.

From Table \ref{Tab2}, 
escapers correspond to roughly 3.3 per cent of all destroyed 
CVs. Most of the escapers escape due to relaxation 
(94 per cent).  They are mostly stable CVs, in general, before 
they escape.  This means that, had they not escaped, 
they would remain as CVs in the cluster.

Unstable CVs, with high-mass donors, tend to have longer periods
than stable CVs. Consequently, they are more likely to suffer a
strong dynamical interaction.  On the other hand, they have short
lives, which means that they do not have much 
time to interact.

\subsubsection{CVs destroyed by Dynamical Interactions}
\label{destroyed_CVs}

Finally, some CVs can literally be destroyed by dynamical
interactions. Since the CVs tend to be very hard binaries, the
probability of such destructive interactions is small \citep{Leigh_2016}.
Indeed, in all six models, only 1.7 per cent of the 
destroyed CVs were destroyed by strong dynamical
interactions. As with the escapers, these were stable CVs
before the interaction occurred.

Above all, only 5 per cent of all destroyed
CVs are stable. This means that once a stable CV is formed,
it tends to live for a long time, unless it escapes from the cluster 
or has its orbital parameters changed by a dynamical interaction.

\subsection{CVs formed from escaping binaries}
\label{ESC}

In this last section, we discuss CVs that
form from escaping binaries, ejected sometime within the 
12 Gyr of cluster evolution.  We stress that
the escaping binaries escape at any time and all
CVs were evolved from the time of escape up
to 12 Gyr using the BSE code.

As expected, no CVs escaped from model K1.
This is easy to understand by considering the strength
of dynamical interactions characteristic of this model. The 
cluster is very 
sparse and the probability of interaction (leading to a 
change in the binary parameters) is small.  Consequently, 
the probability of significant changes occurring 
in the escaping binaries (before the escape) is small.
Since not a single CV is formed through pure
binary stellar evolution in the Kroupa models, we see that model K1 should
produce only a few (if any) CVs from the escaping binaries, 
since the influence of dynamics in this model is weak.

Models K2 and K3 have 12 and 23 CVs formed from escaping
binaries, respectively.  In these models, future 
CVs were created prior to escape and, in a field-like
environment, the CVs were able to form.  Analogously to what
happens for the GC CVs in models K2 and K3, CVs formed
from escaping binaries tend to have high-mass WDs.  This is
the result of strong dynamical interactions, specifically 
exchanges.

Models S1 and S3 have, respectively, 34 and 19
CVs formed from escapers. Note that this number 
of CVs is comparable to the number of CVs present in
the cluster at the present-day (41 in S1, and 
27 in S3). This is easy to understand by considering
the evolution of these clusters (Section \ref{GC-EVOL-cluster}).
Both models are post-core-collapse and are expanding
due to the energy source in their cores (S1 due to
a subsystem of black holes and S3 due to binaries).
This makes the escape rate high, which in turn
reflects in the number of CVs formed from 
escaping binaries.

Interestingly, model S2 produces only 4 CVs out of all the 
escaping binaries. This is again connected to the 
cluster evolution. As already pointed out,
this cluster is approaching core-collapse (which
will take place at $\sim$ 13 Gyr).  Consequently, 
the cluster is not expanding, as in S1 and S3. As such, 
only a few future CVs escape from the cluster.

In models K2 and K3, most of the escaping binaries that later 
become CVs escape from 
the cluster between $\sim$ 6 and $\sim$ 8 Gyr.  
In models S1 and S3, however, the escape time has a roughly
flat distribution, with a peak at $\sim$ 10 Gyr.

In all 6 models, 92 CVs formed from
escaping binaries.  Of those, $\sim$ 62 per cent
of CVs have WDs with masses greater
than 0.5 M$_\odot$. Additionally, $\sim$ 61 per cent
of these CVs are period bouncers, which is much less than
the fraction of period bouncers found in the 
GC CVs ($\sim$ 87 per cent). 

In general, the fraction of escaping binaries that become
CVs is $\sim$ $10^{-4}$.  
Unfortunately, due to the small total number of CVs formed from 
escaping binaries (and the relatively small number of simulations), 
we are unable to say more about them.
It seems that a very small fraction of the
Galactic field CVs in the halo might have their origins 
in GCs.

When more models of the MOCCA-SURVEY
are analyzed, we will better constrain the properties
of CVs formed from escaping binaries, and 
their relation to the Galactic field CVs.

\section{CONCLUSIONS}
\label{conclusion}

In the first paper of this series \citep{Belloni_2016a}, we discussed 
six specific MOCCA models with a focus on the properties of
their present-day CV populations. In this paper, we concentrate instead 
on a discussion of the properties of the progenitor and formation-age
populations.

Our results show good overall agreement with previous investigations,
both with respect to the most common CV formation channels \citep{Ivanova_2006}
and the acceleration/retardation of CV evolution prior to CV formation, by 
(indirectly) affecting the CV progenitor binary \citep{Shara_2006} .

The main results of this paper can be summarized as follows:

\begin{description}

\item[(i)] Dynamics can extend the parameter space applicable to CV progenitors 
(with respect to CVs formed without influence of dynamics), and allow binaries 
that would not become CVs to evolve into CVs.
\item[(ii)] Sparse clusters have more CVs formed through a CEP, relative to 
denser clusters with more dynamically formed CVs. The number of dynamically 
formed CVs decreases with decreasing cluster density, as expected. 
\item[(iii)] The WD-MS binary formation rate is characterized by an initial burst,
followed by a smoothly decreasing rate.  The CV formation rate 
shows the same initial burst, although it is followed by a nearly
constant formation rate.  This is caused by the time-delay between the time of 
WD-MS binary formation and CV formation. This is, in general,
in good agreement with the findings of \citet{Ivanova_2006}, 
who claimed that the relative number of CVs that appear is roughly constant throughout
the cluster evolution.
\item[(iv)] The CV formation rate can either be accelerated or retarded due to
dynamical interactions, which is in good agreement with the results of \citet{Shara_2006}. 
Additionally, the CV is unlikely to change after CV formation, i.e. after CV formation, it is 
improbable that the CV will be affected by dynamical interactions, 
because the CVs are dynamically hard binaries with small interaction cross-sections.
\item[(v)] The CVs are mainly formed as short-period systems, which indicates that
they will be very faint objects by the present-day. 
\item[(vi)] Dynamically formed CVs tend to have massive WDs, in general, due
to exchanges.  This was pointed out by \citet{Ivanova_2006}.
\item[(vii)] The properties of CV populations change with time, which
can lead to confusion upon comparing predicted cluster CVs (older) 
to observed Galactic field CVs (younger).
\item[(viii)] Before the present-day, CVs can be `destroyed' either via unstable mass 
transfer (most of them), dynamical interactions, or cluster escape, in good 
agreement with \citet{Shara_2006} and \citet{Ivanova_2006}.
\item[(ix)] Very few field CVs could have their origins in GCs.

\end{description}

This concludes the first part of our investigation into CV formation in GCs using the 
MOCCA, split between two first papers of which this is the second.  
Thus far, our results have shown good overall agreement with previous
observational and theoretical works.

Future investigations will concentrate on the effects of the 
empirical consequential angular momentum loss prescription \citep{Schreiber_2016}
(which is associated with the mass loss from the system), 
and on the CEP parameters, in deciding the predicted CV properties. 
Special attention will be given to the absence of field-like CVs in 
the Kroupa models.  Equally interesting are the CV siblings (AM CVn and
symbiotic stars), especially upon considering that the first AM CVn in a 
GC might have just been discovered in NGC 1851 \citep{Zurek_2016}. 
We plan to extend the CATUABA code in 
order to include these interesting systems in order to extend our analysis 
to include the entire population of accreting white dwarf binary systems in GCs.  
After CATUABA is complete and automated, we will begin our 
analysis of the models of the MOCCA-SURVEY \citep{Askar_2016b}.

\section*{Acknowledgements}

DB would like to kindly thank Christian Knigge, M{\'o}nica Zorotovic, and 
Matthias Schreiber for useful discussions and suggestions, which
made the quality of the paper increase substantially. We would also like
to thank the referee Matthew Benacquista for his comments that improved this 
work. DB was supported by the CAPES foundation, Brazilian Ministry of Education
through the grant BEX 13514/13-0. MG and AA were supported by
the National Science Centre through the grant DEC-2012/07/B/ST9/04412. 
AA would also like to acknowledge support
from the National Science Centre through the
grant UMO-2015/17/N/ST9/02573 and partial support from Nicolaus
Copernicus Astronomical Centre's grant for young researchers.

\bibliographystyle{mnras}
\bibliography{references}

\bsp

\label{lastpage}

\end{document}